\documentclass[12pt]{article}
\UseRawInputEncoding

\usepackage{authblk}
\usepackage{titlesec}
\titleformat{\section}
{\normalfont\large\bfseries}{\thesection}{1em}{}

\usepackage{amsthm}
\usepackage{amsmath}
\usepackage{amssymb}
\usepackage{bm}
\usepackage{lscape}
\usepackage{subfigure}
\usepackage{epsfig}
\usepackage{multirow}
\usepackage{pdflscape}
\renewcommand\appendix{\par
  \setcounter{section}{0}
  \setcounter{subsection}{0}
  \setcounter{figure}{0}
  \setcounter{table}{0}
  \renewcommand\thesection{Appendix \Alph{section}}
  \renewcommand\thefigure{\Alph{section}\arabic{figure}}
  \renewcommand\thetable{\Alph{section}\arabic{table}}
}

\usepackage[round]{natbib}
\usepackage[margin=1.0in]{geometry}
\usepackage{setspace}
\usepackage{hyperref}
\usepackage{graphicx}
\usepackage{color}
\usepackage{dcolumn}

\usepackage{amsfonts,amscd}
\usepackage{lastpage}
\usepackage{enumerate}
\usepackage{fancyhdr}
\usepackage{mathrsfs}
\usepackage{xcolor}
\usepackage{listings}
\usepackage{etoolbox}
\usepackage{float}

\usepackage{color}
\usepackage{sectsty}
\usepackage{comment}
\usepackage{caption}
\usepackage{pdflscape}
\usepackage{array}
\usepackage{longtable}

\usepackage{multirow}
\usepackage{makecell}

\usepackage[utf8]{inputenc}

\usepackage{booktabs}

\usepackage{calc} 

\sectionfont{\large}

\begin{document}

\begin{titlepage}
	\title{The Marginal Labor Supply Disincentives of Welfare: \\ Evidence from Administrative Barriers to Participation\thanks{The authors would like to thank Marc Chan, Kai Liu, Shaiza Qayyum, Kyungmin Kang, and Sue Bahk for research assistance as well as the participants of a large number of conferences and departmental seminars and numerous specific individuals for comments, including formal discussant remarks by James Ziliak. Comments at a seminar at the University of Chicago were particularly helpful. Comments from the Editor and referees were also valuable in improving the paper.  Research support from the National Institutes of Health is gratefully acknowledged.}}
	\author{Robert A. Moffitt\thanks{Corresponding author: \texttt{\href{mailto:moffitt@jhu.edu}{moffitt@jhu.edu}}.} \\ Johns Hopkins University \and Matthew V. Zahn \\ Johns Hopkins University}
	\date{Original: July 2019 \\ Revised: January 2022}
	\maketitle
	\thispagestyle{empty}
	\clearpage
	\begin{abstract}
		\noindent Existing research on the static effects of the manipulation of welfare program benefit parameters on labor supply has allowed only restrictive forms of heterogeneity in preferences. Yet preference heterogeneity implies that the marginal effects on labor supply of welfare expansions and contractions may differ in different time periods with different populations and which sweep out different portions of the distribution of preferences.  A new examination of the heavily studied AFDC program uses variation in state-level administrative barriers to entering the program in the late 1980s and early 1990s to estimate the marginal labor supply effects of changes in program participation induced by that variation.  The estimates are obtained from a theory-consistent reduced form model which allows for a nonparametric specification of how changes in welfare program participation affect labor supply on the margin.  Estimates using a form of local instrumental variables show that the marginal treatment effects are quadratic, rising and then falling as participation rates rise (i.e., becoming more negative then less negative on hours of work).  The average work disincentive is not large but that masks some margins where effects are close to zero and some which are sizable.  Traditional IV which estimates a weighted average of marginal effects gives a misleading picture of marginal responses.  A counterfactual exercise which applies the estimates to three historical reform periods in 1967, 1981, and 1996 when the program tax rate was significantly altered shows that marginal labor supply responses differed in each period because of differences in the level of participation in the period and the composition of who was on the program. \\
		\vspace{0in}
		\vspace{0in}
		
		\bigskip
	\end{abstract}
	\setcounter{page}{0}
	\thispagestyle{empty}
\end{titlepage}
\pagebreak \newpage

\doublespacing
\setlength{\parindent}{5ex}


The classic form of a welfare program for a low-income population is that represented by a negative income tax, with a guaranteed minimum cash payment for those with no private income and with a positive marginal benefit-reduction rate, or tax rate, applied to increases in earnings.  In the U.S., the only major cash program that has taken this classic shape was the Aid to Families with Dependent Children (AFDC) program, which took that shape from its formation in 1935 to the early 1990s, when its structure was changed.  Notable reforms in the program took place in 1967, 1981, and 1996, with a decrease in the nominal tax rate in the first year from 100 percent to 67 percent, an increase in the tax rate back to 100 percent in the second year, and a decrease in the tax rate again in the third year to approximately 50 percent (albeit accompanied by many other reforms). The effects of these reforms on labor supply have been heavily studied (see \cite{Moff1992,Moff2003}, and \cite{Ziliak2016} for reviews).

This paper revisits this literature, arguing that the empirical models used to evaluate the impact of welfare participation on labor supply have been excessively restrictive in the representation of unobserved heterogeneity in the eligible population (i.e., heterogeneity conditional on the observables). By definition, the effect of any reform on labor supply depends on the labor supply responses of inframarginal individuals (i.e., those who remain on the program both before and after the reform) but also on the labor supply responses of marginal individuals who either join or leave the program in response to the reform.  With sufficient heterogeneity of preferences, these two responses are not the same, but the existing literature on the effects of AFDC reforms on labor supply has almost entirely assumed they are equivalent.  

\begin{figure}[H]
	\centering
	\caption{AFDC-TANF Total Caseload and Per Single-Mother Family, 1967--2015}\label{fig:AFDC_TANFcases}
	\begin{minipage}{\linewidth}
		\includegraphics[width=1.1\textwidth]{"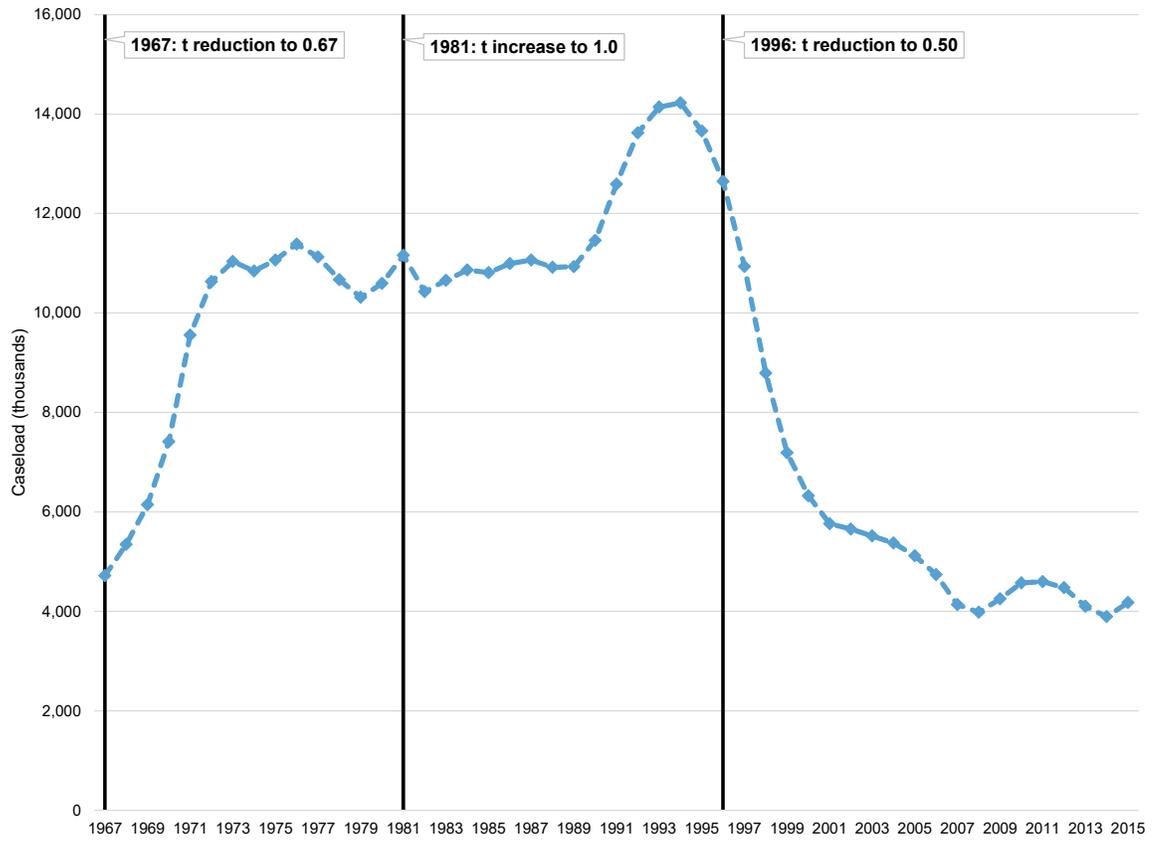"}
		\rule{\linewidth}{0in}
	\end{minipage}
\end{figure}

That the composition of the caseload is likely to have changed over time is suggested by Figure \ref{fig:AFDC_TANFcases}, which shows how the caseload changed over the period 1967--2015.  The caseload rose in the late 1960s, flattened out over the 1970s and the 1980s, rose again in the early 1990s, and fell sharply thereafter. The nominal tax rate in the program was reduced from 1.0 to 0.67 in 1967 and was raised back to 1.0 in 1981. Models with homogeneous preferences would predict that the second tax rate change would simply reverse the former, conditional on changes in other observables.  But if the much higher caseload in 1981 compared to 1967 was associated with a different caseload composition, the marginal individual would have different labor supply responses and hence the two reforms would not have equal and opposite effects. The reduction in the tax rate from 1.0 down to 0.5 could also have had a different response because the caseload was much higher than that in any prior period.

In this study, we estimate marginal labor supply effects of changes in AFDC program participation in the late 1980s and early 1990s---the last period the program took its classic form, as noted above. The paper contains a theoretical exposition of marginal labor supply effects in the textbook static labor supply model; proposes a reduced form econometric model designed to be estimated with local instrumental variables; and proposes instruments draw from variation in administrative barriers arising from errors made by states in allowing applicants onto the program.

The first section lays out the familiar static labor supply model in the presence of a classic welfare program but adds two sources of heterogeneity: labor supply preference heterogeneity and heterogeneity in the fixed costs of participation. This leads to a formal definition of marginal individuals as those who lie on a locus defined by the values of those two parameters which put the individual on the margin of participation. Participation and non-participation are then defined by whether individuals' two parameters lie on either side of the locus. As program parameters change, or as the distribution of fixed costs changes, the locus shifts and individuals either enter or leave the program.  The mean labor supply responses of those who change participation define the marginal labor supply response (or marginal treatment effect).  It is also shown that, as the program is continually expanded or contracted, marginal responses can grow, fall, or remain the same in an arbitrary and unrestricted fashion.

The second section presents a reduced form model designed for the estimation of marginal labor supply responses.   Following the original proposal for marginal treatment effects (\cite{BjorkMoff1987}), it is set up as a random coefficients model.\footnote{\cite{HeckRobb1985} earlier introduced the random coefficients model to represent unobserved heterogeneity in treatment effects.}  The random coefficients formulation is in all respects equivalent to the conventional causal model but makes marginal treatment effects more transparent. Using a modified form of the local instrumental variable estimation outlined by \cite{HeckVyt1999,HeckVyt2001,HeckVyt2005,HeckVyt2007} in their extensive development of the marginal treatment effects model, an approach is developed to estimate the marginal treatment effect curve over the support of the participation rate (i.e., the propensity score) nonparametrically with sieve methods. The parameters of the reduced form model are directly related to those of the structural model and are fully theory-consistent.

The third section estimates the form model with cross-sectional data from 1988 to 1992 from the Survey of Income and Program Participation.  The instruments used are measures of administrative barriers imposed by states in handling applications to the program, and cross-state variation in those measures are shown to affect participation rates in the state. The instrument is shown to be weak in some regions of the propensity score but strong in others, so only estimates in the strong regions are considered.  The estimates show that marginal labor supply responses are quadratic, growing more negative as participation expands and then turning less negative and eventually statistically insignificant as participation expands beyond a certain level. While the average marginal response across all margins is not large, this masks some margins along which responses are close to zero and others where they are sizable. It is also shown that the quadratic relationship is explained by a pattern of responses of full-time and part-time workers that changes as participation expands. 

The fourth section reports robustness checks for the validity of the instruments, in one case using regression discontinuity methods based on close elections and, in the other case, using Congressional legislation in 1988 that caused states to differentially change their error rates starting in 1992.  The checks show the same pattern of marginal labor supply responses as in the main analysis. Falsification tests are also reported which show the instruments to have no effects on labor supply among demographic groups ineligible for the program.

The fifth and final section conducts a simple counterfactual exercise which asks who was on the margin at the time of the major 1967, 1981, and 1996 reforms. The exercise assumes the model estimated over the 1988--1992 period would still apply but that guarantees, tax rates, demographics, and the level of participation instead took on their values in those three years. Marginal responses are calculated to be the largest in 1967, the smallest in 1981, and in between in 1996.  The calculations also show that the reduction in the tax rate from 100 percent to 67 percent in 1967 and the increase in the tax rate back to 100 percent in 1981 did not have symmetric effects because the populations on the margin were different in those two years.\footnote{Fortunately, the support of the participation rates used in the empirical analysis (where the instruments are strong) include the participation rates in those three years. Hence methods for extrapolation beyond the support of participation rates in the data, as discussed by, for example, \cite{BrinMogWis2017} and \cite{MogSanTor2018}, are not needed.}

The paper brings together two literatures.  One is the massive literature on the labor supply effects of welfare guarantees and tax rates, and the effects of various welfare reforms, on labor supply.  References to reviews of that literature were given above.  That literature typically estimates the effects of program parameters and reforms on labor supply, which cannot identify labor supply effects of those on the margin because inframarginal recipients are also affected by those variables.  The second is the literature on marginal treatment effects begun by \cite{BjorkMoff1987} and extensively developed by \cite{HeckVyt1999,HeckVyt2001,HeckVyt2005,HeckVyt2007}, as noted above. Both important theoretical and empirical work on marginal treatment effects has been conducted since that time, but the majority of the empirical work has been on marginal treatment effects on earnings from additional schooling (\cite{CarnHeckVyt2010,CarnHeckVyt2011}, and \cite{HeckUrzuaVyt2006}).  Applications of marginal treatment estimation to other areas include studies of foster care and child removal (\cite{Doyle2007} and \cite{Baldetal2019}), the Social Security Disability Insurance program (\cite{MaeMulStr2013}), health insurance (\cite{Kow2016}), early child care (\cite{Cornetal2018}), and incarceration (\cite{Bhuletal2020}). This paper is the first to apply the method to the effect of welfare programs on labor supply.


\singlespacing\section{Adding Heterogeneity to the Canonical Static Labor Supply Model of Transfers}
\doublespacing
\hspace{5ex}The canonical static model of the labor supply response to transfers (\cite{Moff1983}, and \cite{ChanMoff2018}) assumes utility to be
\begin{equation} \label{eq:1}
U(H_{i},Y_{i};\theta _{i})-\phi _{i}P_{i}  
\end{equation}
where $H_{i}$ is hours of work for individual $i$, $Y_{i}$ is disposable income, $P_{i}$ is a program participation indicator, $\theta _{i}$ is a vector of labor supply preference parameters, and $\phi _{i}$ is a scalar representing fixed costs of participation in utility units whose distribution is in the positive domain.  The presence of $P_{i}$ allows for the presence of fixed costs of participation---in money, time, or utility (stigma), with the exact type unspecified and scaled in units of utility (\cite{Moff1983}, \cite{DaponteSanTay1999}, and \cite{Currie2006}).  Some type of cost is required to fit the data on almost all transfer programs because many individuals who are eligible for transfer programs do not participate in them, and some type of cost is the usual explanation for that feature of the data.  Separability of the fixed costs from the utility of leisure and income is not necessary for the theoretical model but is required for the econometric model, as we discuss below, so that separability is maintained at the outset.\footnote{The existence of a cost function also opens an avenue for instruments that affect fixed costs but not hours of work directly, the same role that cost functions often play in models of schooling and human capital (see e.g., p.674 of \cite{HeckVyt2005}). This will be the source of the instrument in the empirical work in this paper.}

\begin{figure}[H]
	\centering
	\caption{Traditional Income-Leisure Diagram}\label{fig:incLeisure}
	\begin{minipage}{\linewidth}
		\includegraphics{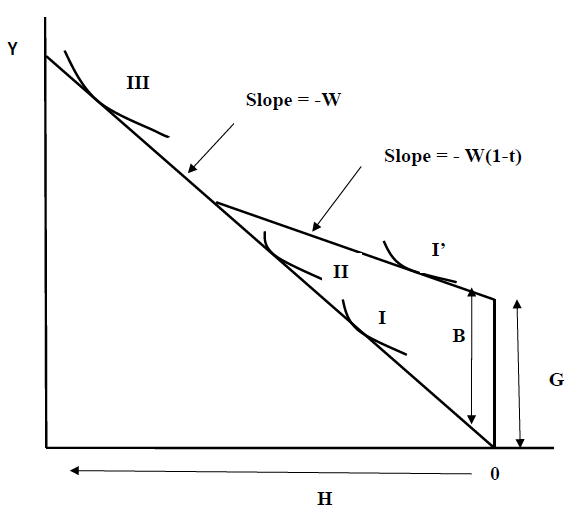}
		\rule{\linewidth}{0in}
	\end{minipage}
\end{figure}

The individual faces an hourly wage rate $W_{i}$ and has available exogenous non-transfer nonlabor income $N_{i}$.  The welfare benefit formula is $
B_{i}=G-tW_{i}H_{i}-rN_{i}$ (assuming, for the moment, that the parameters $G$, $t$ and $r$ do not vary by $i$) and hence the budget constraint is
\begin{align}
	Y_i = \begin{cases}
		 	W_{i}(1-t)H_{i}+G+(1-r)N_{i} &\text{if $P_i =1$} \\ 
		 	W_{i}H_{i}+N_{i} &\text{if $P_i =0$}
		  \end{cases}
\end{align}
The resulting labor supply model is represented by two functions, a labor supply function conditional on participation and a participation function:
\begin{equation} \label{eq:3}
H_{i}=H[W_{i}(1-tP_{i}),N_{i}+P_{i}(G-rN_{i});\theta _{i}]  
\end{equation}
\begin{equation} \label{eq:4}
P_{i}^{\ast }=V[W_{i}(1-t),G+N_{i}(1-r);\theta _{i}]-V[W_{i},N_{i};\theta
_{i}]-\phi _{i}  
\end{equation}
\begin{equation} \label{eq:5}
P_{i}=1(P_{i}^{\ast }\geq 0)  
\end{equation}%
where $H$ is the labor supply function, $V$ is the indirect utility function and 1($\cdot $) is the indicator function. Nonparticipants, those for whom $P^*$ is negative, are of two types: low-work individuals for whom a positive benefit is offered and a utility gain (in $V$) could be obtained but who do not participate because $\phi _{i}$ is too high, and high-work individuals for whom the utility gain (in $V$) is negative and who would not participate even if $\phi _{i}$ were zero (these individuals are above the eligibility point). Figure \ref{fig:incLeisure} is the familiar income-leisure diagram showing three different individuals who respond to the transfer program constraint by continuing to work above the eligibility point (III), working below that point but off the program (II), and working below that point but on the program (I'; I is the pre-program location for this individual).

Equations \eqref{eq:3}--\eqref{eq:5} are in the form of a generalized Roy model, but where the outcomes for the two regimes are notationally represented in the single equation \eqref{eq:3} instead of two separate equations.  The fixed cost term $\phi_{i}$ plays the role of the cost term in the generalized Roy model while the change in $V$ corresponds to the gain in earnings or other outcome in that model.  Unlike the Roy model where the earnings gain is typically assumed to be linear in the selection equation (e.g., \cite{HeckVyt2005}), here the unobservable $\theta_{i}$ enters nonlinearly through the indirect utility function $V$.  Consequently, the selection equation does not have a composite error term which is a linear combination of the component errors. 

The labor supply response to the program for individual $i$ conditional on the budget constraint parameters is the change in hours worked when participating: 
\begin{equation} \label{eq:6}
\bigtriangleup _{i}(\theta _{i}| C_{i})=H[W_{i}(1-t),G+N_{i}(1-r);\theta_{i}]-H[W_{i},N_{i};\theta _{i}]
\end{equation}
where $C_{i}=[W_{i},N_{i},G,t,r]$ is the set of budget constraint variables.  The response in equation \eqref{eq:6} is a heterogeneous response if $\theta _{i}$ varies with $i$. There is a latent distribution of these responses for the full population, including those who do not eventually participate. 

To define the marginal labor supply response, or marginal treatment effect, first note that equation \eqref{eq:4} implies that individuals on welfare must have increases in $V$ from participation that are greater than their $\phi_{i}$ values.  A reduction in fixed costs represented by a downward shift in the distribution of the $\phi_{i}$ will bring onto the program those whose increases in $V$ had put them just on the margin of participation initially.  The values of $\bigtriangleup_{i}$ for those individuals are the labor supply responses of those on the margin.

More formally, define $\theta_{D}$ and $\phi_{D}$ as the values which make an individual indifferent between participation and non-participation:
\begin{equation} \label{eq:8}
\begin{aligned}
0&=V[W_{i}(1-t),G+N_{i}(1-r);\theta _{D}]-V[W_{i},N_{i};\theta_{D}]-\phi _{D}  \\
&=dV(\theta_{D}|C_{i})-\phi_{D}
\end{aligned}
\end{equation}
where the second line just defines $dV(\theta_{D}|C_{i})$.  Equation \eqref{eq:8} defines a locus of the two unobservables along which marginal individuals locate.  That locus is shifted by the budget constraint parameters.  Following the literature (e.g., \cite{HeckVyt2005}, equation (4)), the marginal labor supply response can be defined as $\bigtriangleup^{MTE}(C_{i})=E_{\phi_{D}}\bigtriangleup[\theta_{D}(\phi_{D},C_{i})|C{i}]$ where $\theta_{D}(\phi_{D},C_{i})$ is the function solving equation \eqref{eq:8} for $\theta_{D}$ as a function of $\phi_{D}$ and $C_{i}$.\footnote{As previously noted, in the typical generalized Roy model, the unobservables are linearly related in the indifference locus and hence only the composite error term matters for selection.  Here, with the unobservables nonlinearly related, selection depends on the two unobservables separately.}

A question is whether the labor supply responses of those on the margin are greater or smaller than those initially on the program, holding constant $C_{i}$.  The answer is that the sign is ambiguous.  While those on the margin have, by definition, smaller values of $dV(\theta_{i}|C_{i})$ than those initially on the program holding fixed costs constant, there is no necessary relationship between the magnitude of those utility differences and the magnitudes of the $\bigtriangleup_{i}|C_{i}$.  Intuitively, the utility gain $dV(\theta_{i}|C_{i})$ is achieved by some combination of an increase in leisure and an increase in goods consumption. The mix depends on relative preferences for those two goods, and those relative preferences can vary arbitrarily over the $dV$ distribution.  Consequently, for example, as $\phi_{i}$ falls in successive increments and as program participation rises, $\bigtriangleup_{i}$ can rise, fall, or remain the same in any arbitrary pattern.\footnote{As in the generalized Roy model, there is positive selection on gains to participation conditional on costs, but positive selection occurs on $V$, not $\bigtriangleup$, and those two variables do not have a monotonic  relationship.} It is the goal of the empirical work in the sections below to identify that pattern.

\begin{figure}[H]
	\centering
	\caption{Hypothetical dV curve showing regions of $\bigtriangleup$ Among Participants (1,2,3)}\label{fig:HypodV}
	\begin{minipage}{\linewidth}
		\includegraphics{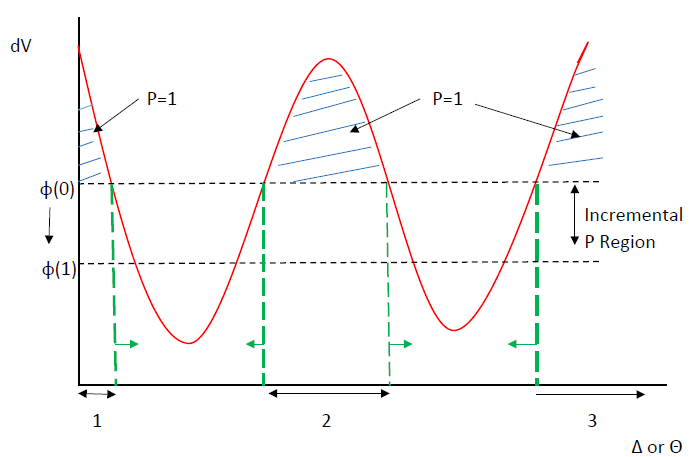}
		\rule{\linewidth}{0in}
	\end{minipage}
\end{figure}

Two figures illustrate these points.  In Figure \ref{fig:HypodV}, a hypothetical pattern of a relationship between $dV_{i}$ and $\bigtriangleup_{i}$ is shown.  While $dV$ is a function of $\theta$ and not $\bigtriangleup$, $\theta$ can be defined without loss of generality to be monotonically related to $\bigtriangleup$ and hence the horizontal axis can be represented with either parameter.  Figure \ref{fig:HypodV}, reflecting the just-mentioned result that $dV$ and $\bigtriangleup$ can have any arbitrary relationship, assumes that they have an alternating pattern of positive and negative association.  For individuals with a value of $\phi_{0}$, three regions are identified where $P=1$ and each is associated with a range of labor supply responses, $\bigtriangleup$ (those ranges are labeled 1, 2, and 3 on the horizontal axis).  A fall in the value of $\phi$ to $\phi_{1}$ increases participation, and the regions of $\bigtriangleup$ of participation expand. The mean $\bigtriangleup$ of those newly joining the program is the integral over the distribution of $\bigtriangleup$ in the new regions of participation.  Of course, in actuality there is a joint distribution of $\bigtriangleup$ and $\phi$, so the actual regions of participation and of $\bigtriangleup$ will depend on that joint distribution and must be integrated over both.

\begin{figure}[H]
	\centering
	\caption{Hypothetical $\Theta_D$, $\phi_D$ Locus}\label{fig:HypoTp}
	\begin{minipage}{\linewidth}
		\includegraphics{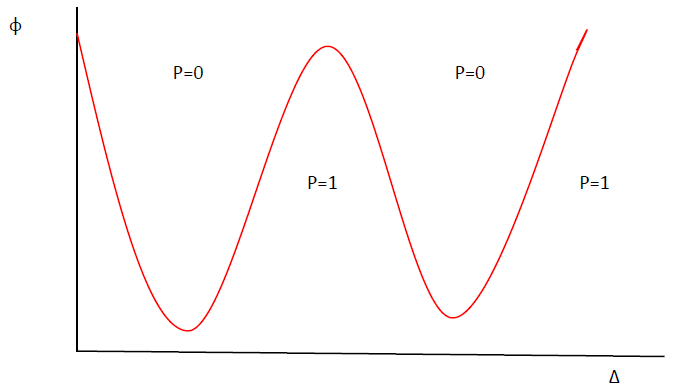}
		\rule{\linewidth}{0in}
	\end{minipage}
\end{figure}

Since the locus of indifference is where $dV=\phi$, the indifference locus showing the values of $\phi$ which make participation marginal for any value of $\theta$ or $\bigtriangleup$---that is, the locus corresponding to equation \eqref{eq:8}---will have the same pattern as Figure \ref{fig:HypodV}.  It is shown in Figure \ref{fig:HypoTp}, along with the regions where $P=0$ and $P=1$. The joint distribution of the two parameters determines the magnitude of the participation and non-participation rates. The locus is shifted when the budget constraint parameters $C_{i}$ change or when the parameters of the joint distribution of $\theta_{i}$ and $\phi_{i}$ shift.  For example, if $\phi_{i}=\overline \phi(Z_{i})+\nu_{i}$, where $Z_{i}$ is an observable proxy for costs and $\nu_{i}$ represents unobserved costs, the line of indifference is the same as in Figure \ref{fig:HypoTp} but with the vertical axis measuring $\nu$ instead of $\phi$, and with the indifference line understood to be conditional on $Z_{i}$. A shift in $Z_{i}$ hence shifts the indifference locus.

The marginal labor supply response like that illustrated in Figure \ref{fig:HypodV} is typically identified by a change in the mean effect of the treatment on the treated (i.e., the mean labor supply response of participants) as participation expands. That mean in this model is
\begin{equation} \label{eq:9}
\begin{aligned}
\widetilde{\bigtriangleup }_{P_{i}=1}&=E(\bigtriangleup _{i}|C_{i},P_{i}=1) \\
&=\frac{1}{P} \int\limits_{S_{\theta \phi }}^{{}}\int\bigtriangleup _{i}(\theta _{i}|C_{i})dJ(\theta _{i},\phi_{i})
\end{aligned}
\end{equation}
where $S_{\theta \phi }$ is the set of parameters in regions demarcated by the $\theta_{D}$,$\phi_{D}$ locus which generates $P=1$, where $J(\theta_{i},\phi_{i})$ is the joint distribution function of $\theta_{i}$ and $\phi_{i}$, and where
\begin{equation} \label{eq:7}
\begin{aligned}
P&=E(P_{i}|C_{i}) \\
&=\int\limits_{S_{\phi }}\int\limits_{S_{\theta}}1\{V[W_{i}(1-t),G+N_{i}(1-r);\theta _{i}]-V[W_{i},N_{i};\theta _{i}]-\phi_{i}\}dJ(\theta _{i},\phi _{i})  
\end{aligned}
\end{equation}
is the participation rate ($S_{\theta }$ and $S_{\phi }$ represent the unconditional supports of the two parameters).  The mean effect of the transfer program over the entire population, participants and non-participants combined, conditional on the budget constraint, is
\begin{equation} \label{eq:10}
\begin{aligned}
\widetilde{\bigtriangleup }&=E(\bigtriangleup _{i}P_{i}|C_{i}) \\
&=\int\limits_{S_{\theta \phi }}^{{}}\int\bigtriangleup _{i}(\theta _{i}|C_{i})dJ(\theta _{i},\phi_{i})  
\end{aligned}
\end{equation}

The marginal treatment effect is traditionally defined as the marginal response to an exogenous increase in program participation, which in the notation here is the mean $\bigtriangleup $ of those who change participation, or $\partial \widetilde{\bigtriangleup }$/$\partial P$.\footnote{The MTE is more usually defined as the derivative of $E(H_{i}|P_{i})$ w.r.t. $P_{i}$ (ignoring other conditioning covariates) but since $E(H_{i}|P_{i})=constant+\widetilde{\bigtriangleup }$ (see next section), the two are equivalent.  This formulation of the MTE is often described in estimation terms, as the LIV estimator; see equation (7) in \cite{HeckVyt2005}.  This will be the formulation in the econometric model below, in equation \eqref{eq:11}.}  The values of the response quantities $\bigtriangleup _{i}$, $\widetilde{\bigtriangleup }$, $\widetilde{\bigtriangleup }_{P_{i}=1}$, and $\partial \widetilde{\bigtriangleup } /\partial P$ must all be nonpositive according to theory.


\section{A Reduced Form Econometric Model}

\hspace{5ex}The objective of the empirical work is to estimate the marginal effect on hours of work of a change in participation induced by a change in fixed costs. Equation \eqref{eq:3} implies that, definitionally,
\begin{equation} \label{eq:11}
\begin{aligned}
H_{i}&=P_{i}H[W_{i}(1-t),G+(1-r)N_{i};\theta_{i}]+(1-P_{i})H(W_{i},N_{i};\theta_{i}) \\
&=H(W_{i},N_{i};\theta_{i})+P_{i}\bigtriangleup _{i}
\end{aligned}
\end{equation}
where $\bigtriangleup _{i}$ is defined in equation \eqref{eq:6}. Now assume that $\phi_{i}=m(Z_{i})+\nu_{i}$, where $Z_{i}$ is an observable correlate of fixed costs and $\nu_{i}$ represents variation in $\phi_{i}$ conditional on $Z_{i}$. Then mean hours of work in the population conditional on the budget constraint and on $Z_{i}$ can be expressed as
\begin{multline} \label{eq:12}
E(H|W,N,G,t,r,Z)=E_{\theta }[H(W,N;\theta )\mid W,N] \\
+E_{\theta ,\nu}(\bigtriangleup \mid P=1,W,N,G,t,r,Z)E_{\theta,\nu}(P|W,N,G,t,r,Z)
\end{multline}
where individual subscripts have been omitted for simplicity.  Both the left hand side and the last term on the RHS are identified in the data so the question is whether the conditional mean of $\bigtriangleup$ can be (this is the effect of the treatment on the treated and was expressed in the last section as equation \eqref{eq:9}). This can be most easily seen, and the estimation method also clarified, by first implicitly conditioning on the budget constraint and all other $X$ variables so as not to have to carry along their conditioning explicitly.  Then equation \eqref{eq:11} (which comes from equation \eqref{eq:3}) and its associated equations \eqref{eq:4}--\eqref{eq:5} can be written as
\begin{equation} \label{eq:14}
H_{i}=\beta _{i}+\alpha _{i}P_{i} 
\end{equation}%
\begin{equation} \label{eq:15}
P_{i}^{\ast }=m(Z_{i})+\delta _{i}
\end{equation}%
\begin{equation} \label{eq:16}
P_{i}=1(P_{i}^{\ast }\geq 0)  
\end{equation}%
where $\beta _{i}$ is hours worked off welfare and $\alpha _{i}$ is a relabeling of $\bigtriangleup_{i}$, the effect on hours of work from going onto welfare for individual $i$. Equation \eqref{eq:14} is equivalent to equation \eqref{eq:3} and the $\beta_{i}$ and $\alpha_{i}$ constitute the elements of $\theta_{i}$ in this formulation. Equation \eqref{eq:14} is also equivalent to the conventional two-regime model with separate outcomes if treated and not treated because $H_{i}=\alpha_{i}+\beta_{i}$ if $P_{i}=1$ and $H_{i}=\alpha_{i}$ if $P_{i}=0$.  The equivalent random coefficient formulation is used here because it will be the basis for the estimating equation.  The participation equation in equations \eqref{eq:15}--\eqref{eq:16} is a representation of equations \eqref{eq:4}--\eqref{eq:5} and the parameter $\delta_{i}$ combines the three parameters $\alpha_{i}$, $\beta_{i}$, and $\nu_{i}$.  Those three parameters are allowed to be individual-specific and to have some unrestricted joint distribution.\footnote{This model is equivalent to that in equations (2a)--(3) in \cite{HeckVyt2005}, and the parameter $\delta$ is equivalent to the well-known $U_{D}$ in that model.}

The object of interest is the distribution of $\alpha_{i}$. Selection in this model can occur either on the intercept ($\beta _{i}$) or the slope coefficient ($\alpha _{i}$) or both because both may be related to $\delta_{i}$ and, in fact, they must be because $\delta_{i}$ contains $\alpha_{i}$ and $\beta_{i}$.  Equation \eqref{eq:12}, which conditions on $Z_{i}$ and hence is a reduced form, and the associated participation equation, now take the form
\begin{equation} \label{eq:17}
E(H_{i}\mid Z_{i}=z)=E(\beta _{i}\mid Z_{i}=z)+E(\alpha _{i}\mid P_{i}=1,Z_{i}=z)\Pr (P_{i}=1\mid Z_{i}=z)  
\end{equation}
\begin{equation} \label{eq:18}
E(P_{i}\mid Z_{i}=z)=\Pr [\delta _{i} \geq -m(z)]  
\end{equation}
Identification of $E(\alpha _{i}\mid P_{i}=1,Z_{i}=z)$ requires, at minimum, that $Z_{i}$ satisfy two mean independence requirements, one for the intercept and one for the slope coefficient:
\begin{equation*} \label{eq:19}
\hspace{0.1in}E(\beta _{i}\mid Z_{i}=z)=\beta\tag{A1} 
\end{equation*}
\begin{equation*} \label{eq:20}
\hspace{0.1in}E(\alpha _{i}\mid P_{i}=1,Z_{i}=z)=g[E(P_{i}\mid Z_{i}=z)]\tag{A2}
\end{equation*}
where $g$ is the effect of the treatment on the treated conditional on $Z_{i}$.  That effect depends on the shape of the distribution of $\alpha _{i}$ and how different fractions of participants are selected from different portions of that distribution. While \eqref{eq:19} is familiar, \eqref{eq:20} may be less so. The usual assumption in the literature is that the two potential outcomes, $\beta_{i}$ and $\beta_{i}+\alpha_{i}$, are fully independent of $Z_{i},$ which implies that $\alpha_{i}$ is as well. Equation \eqref{eq:20} is a slightly weaker condition which states that all that is required is that the mean of $\alpha_{i}$ conditional on participation be independent of $Z_{i}$ conditional on the participation probability (i.e., the propensity score). Variation in $Z_{i}$ generates variation in participation which generates variation in the conditional mean of $\alpha_{i}$, but there should be no other channel by which $Z_{i}$ affects that conditional mean.\footnote{The terms ``propensity score" and ``participation probability" are used interchangeably throughout.}

Inserting \eqref{eq:19} and \eqref{eq:20} into the main model in equations \eqref{eq:17}--\eqref{eq:18}, and
denoting the participation probability as $F(Z_{i})=E(P_{i}\mid Z_{i})$, we obtain two estimating equations
\begin{equation} \label{eq:22}
H_{i}=\beta + g[F(Z_{i})]F(Z_{i})+\epsilon_{i}  
\end{equation}
\begin{equation}\label{eq:23}
P_{i}=F(Z_{i})+\xi_{i}  
\end{equation}
where $\epsilon_{i}$ and $\xi_{i}$ are mean zero and orthogonal to the RHS by construction. No other restriction on these error terms need be made, as this is a reduced form of the model. 

Equation \eqref{eq:22} is the key to the estimation approach taken here and will be used for that estimation. It shows that the population mean of $H_{i}$ (that is, taken over participants and nonparticipants) equals a constant plus the mean response of those in the program times the fraction that is in it. The implication of this way of specifying the model---that is, as a random coefficient model---is that preference heterogeneity is detectable by a nonlinearity in the response of the population mean of $H_{i}$ to changes in the participation probability. If responses are homogeneous and hence the same for all members of the population, the function $g$ reduces to a constant and therefore a shift in the fraction on the program has a linear effect on the population mean of $H_{i}$. However, if the responses of those on the margin vary, the response of the population mean of $H_{i}$ to a change in participation will depart from linearity.\footnote{This point is also already in \cite{HeckVyt2005}.  See the discussion on pp.690--692 and, particularly, Figure 2A.}

Equation \eqref{eq:22} appears in the derivative of equation (7) in \cite{HeckVyt2005} in the definition of the LIV estimator. The only difference is that, in that study, $g[F(Z_{i})]F(Z_{i})$ is collapsed into a single function of $F(Z_{i})$ and LIV estimation is conducted by a direct nonparametric computation of the slope of the outcome-propensity-score regression line. Equation \eqref{eq:22} just factors $F(Z_{i})$ out and labels its coefficient as $g[F(Z_{i})]$, and LIV estimation will proceed allowing that function to be nonparametric in the score. Testing for homogeneity in equation \eqref{eq:22} just requires testing for whether $g$ varies with the score instead of testing, equivalently, whether the outcome is quadratic in the propensity score.

The separability of fixed costs from the gain in indirect utility $V$ in the theoretical model (at least from its $\theta_{i}$ component)---see equation\eqref{eq:1}---and the separability of $m(Z_{i})$ and $\nu_{i}$ in the $\phi_{i}$ fixed cost term, is critical to identifying the MTE, as shown by \cite{Vyt2002} and as heavily emphasized by \cite{HeckVyt2005} (see their Section 6). Failure of separability results in failure of index sufficiency and in potential violation of the monotonicity condition (or the ``uniformity" condition in \cite{HeckVyt2005}) in \cite{ImbensAngrist1994} which is needed to guarantee that LIV identifies the MTE.  As in prior work applying MTE methods (\cite{CarnHeckVyt2011}; \cite{MaeMulStr2013}; \cite{Cornetal2018}; and \cite{Bhuletal2020}), we assume separability in our econometric model.  In our discussion of our instruments in the next section, we will argue that our instruments likely satisfy monotonicity as well; we postpone a discussion until that section.

When nonparametric identification of the parameters of the model---$\beta $, the function $g$ at every point $F$, and $F$ itself---is possible has been extensively discussed in the literature and need only be briefly stated.  $F$ is identified at every data point $Z_{i}$ from the second equation from the mean of $P_{i}$ at each value of $Z_{i}$ (apart from sampling error). With identification of $F$, the LATE of \cite{ImbensAngrist1994} is identified by the discrete difference in $H$ between two points $z_{i}$ and $z_{j}$  divided by the difference in $F$ between those two points. With multiple values of $z$, multiple LATE values can be identified.  A marginal treatment effect is a continuous version of this and requires some smoothing method across discrete values of $Z$, and is computed by $\partial H/\partial F=g^{\prime }(F)F+g(F)$.  However, while the MTE $\partial H/\partial F$ is identified, $g$ and $g^{\prime}$ are not unless there is a value of $Z_{i}$ in the data for which $F(Z_{i})=0$.  In that case, $\beta $ is identified from the mean of $H_{i}$ at that point and hence $g$ is identified pointwise at every other value of $z$ since $F$ is identified. If no such value is in the data, then $g$ can only be identified subject to a normalization of its value at a particular value of $z$ or if the value of $g$ is known at some value.

To generate actual estimating equations, we now reintroduce the budget constraint parameters and other $X$ variables and express the reduced form by conditioning on those quantities as well as on $Z_{i}$, leading to equation \eqref{eq:12} with the identifying restrictions imposed:
\begin{equation} \label{eq:230}
\begin{aligned}
E(H\mid W_{i},N_{i},G,t,r,Z_{i})&=E_{\theta }[H(W_{i},N_{i})]+E_{\theta ,\nu}[\bigtriangleup _{i}\mid W_{i},N_{i},G,t,r,P_{i}=1]E_{\theta ,\nu }(P_{i}\mid W_{i},N_{i},G,t,r,Z_{i}) \\
&=h_{0}(W_{i},N_{i})+g[W_{i},N_{i},G,t,r,F(W_{i},N_{i},G,t,r,Z_{i})] F(W_{i},N_{i},G,t,r,Z_{i}) 
\end{aligned}
\end{equation}

As just noted, the intercept $h_{0}(W_{i},N_{i})$ cannot be identified nonparametrically without a value of $F(Z)=0$ in the data and the function $g$ cannot be identified without the same value present and even then only at the values of $Z_{i}$ in the data.\footnote{As described below, our data do not contain values of $Z$ which generate $F=0$, so we will not be able to identify the intercept nonparametrically.}

The theory imposes two restrictions on the form of the equation.  First, the intercept of the equation, denoted by the $h_{0}$ function, must not include the welfare program parameters $G$, $t$, and $r$ because the intercept represents labor supply off welfare.  Hence these parameters should not be ``controlled for" in the $H$ regression.  Second, the function $g$, which is the mean labor reduction for those participating in the program, must contain the budget constraint parameters because those parameters affect the labor supply of inframarginal participants.  They must be included so that changes in the coefficient $g$ induced by changes in $F$ hold the budget constraint fixed, which is required for changes in that coefficient with respect to participation to identify the responses only of marginal participants and not those who are inframarginal. Of course, a fully parametric model which makes use of a specific parametric utility function and assumptions on which parameters of that function are heterogeneous would result in specific functional forms for $h_{0}$, $g$, and $F$.

Full nonparametric estimation of the three functions $h_{0}$, $g$, and $F$ would make the estimation subject to the curse of dimensionality.  Considerable dimension reduction can be achieved by using traditional linear indices in the observables, with
\begin{equation} \label{eq:24}
H_{i}=X_{i}^{\beta }\beta +[X_{i}\lambda +\tilde{g}(F(X_{i}\eta +\delta
Z_{i}))]F(X_{i}\eta +\delta Z_{i})+\epsilon _{i}  
\end{equation}%
\begin{equation} \label{eq:25}
P_{i}=F(X_{i}\eta +\delta Z_{i})+\nu _{i}  
\end{equation}%
where $X_{i}^{\beta }$ denotes a vector of exogenous socioeconomic characteristics plus $W_{i}$ and $N_{i}$ and $X_{i}$ denotes a vector which augments $X_{i}^{\beta }$ with the welfare-program variables $G$, $t$, and $r$ ($\tilde{g}$ is the now the conditional mean of $\alpha_{i}$ also conditioned on $X_{i}\lambda$).\footnote{$X_{i}\lambda$ will be normalized to have mean zero to allow the $\tilde{g}$ function to have an intercept. Some specifications to be estimated will interact $X$ with $\tilde{g}$. We thank a referee for noting that, without such interactions, the additivity of $X_{i}\lambda$ and $\tilde{g}$ inside the brackets may provide some identification by itself.} Exogenous characteristics thus linearly affect labor supply off welfare and linearly affect the arguments of the $\tilde{g}$ and $F$ functions.\footnote{We note that the parametric form of the index function inside $F$ now allows $\beta$ to be identified by extrapolation outside the data. However, our focus will only be on the $\tilde{g}$ function and that will not be extrapolated.}  However, the $\tilde{g}$ function will continue to be nonparametrically estimated, using sieve methods (see below; normality will be assumed for $F$, however). With these two functions specified, we will employ two-step estimation of the model, with a first-stage probit estimation of equation \eqref{eq:25} and second-stage nonlinear least squares estimation of equation \eqref{eq:24} using fitted values of $F$ from the first stage. Consistency and asymptotic normality of two-step estimation of nonlinear conditional mean functions with estimated first-stage parameters is demonstrated in \cite{NewMcFad1994}. Standard errors are obtained by jointly block bootstrapping equations \eqref{eq:24} and \eqref{eq:25} at the state level.\footnote{As part of our first stage, we also estimate a wage equation, which is included in the bootstrap procedure.}


\section{Data, Instruments, and Main Results}

\subsection{Data}
\hspace{5ex}The Aid to Families with Dependent Children (AFDC) program is the only major cash welfare program the U.S. has had, at least for the nonelderly and nondisabled, with a structure close to that of the classic form outlined above. It was created the Social Security Act of 1935 and eligibility required the presence of children and the absence of one parent, with the practical implication that the  caseload was almost entirely composed of single women with children.  However, major structural reforms of the program began in 1993 with the introduction of work requirements and time limits, and it has not returned to its classic form since that time.  Consequently, the analysis here will use data on disadvantaged single women with children from the late 1980s to the early 1990s, just before the change in structure occurred.

Suitable data from that period are available from the Survey of Income and Program Participation (SIPP), a household survey representative of the U.S. population which began in 1984 for which a set of rolling, short (12 to 48 month) panels are available throughout the 1980s and 1990s.  The SIPP is commonly used for the study of transfer programs because respondents were interviewed three times a year and their hours of work, wage rates, and welfare participation were collected monthly within the year, making them more accurate than the annual retrospective time frames used in most household surveys.  The SIPP questionnaire also provided detailed questions on the receipt of transfer programs, a significant focus of the survey reflected in its name.  We use all waves of panels interviewed in the Spring of each year from 1988--1992 (only Spring to avoid seasonal variation) and pool them into one sample, excluding overlapping observations by including only the first interview when the person appears to avoid dependent observations.

Eligibility for AFDC in this period required sufficiently low assets and income and, for the most part, required that eligible families be single mothers with at least one child under 18.  The sample is therefore restricted to such families, similar to the practice in past AFDC research.  To concentrate on the AFDC-eligible population, we restrict the sample to those with completed education of 12 years or less, nontransfer nonlabor income less than \$1,000 per month, and between the ages of 20 and 55.  The resulting data set has 3,381 observations.

The means of the variables used are shown in Appendix Table \ref{tab:VariableMeans}. The variables include hours worked per week in the month prior to interview ($H$) (including zeroes), whether the mother was on AFDC at any time in the prior month ($P$), and covariates for education, age, race, and family structure (the state unemployment rate is also used as a conditioning variable).\footnote{The empirical work will report some estimates separating the extensive margin from the intensive margin of $H$.} Thirty-seven percent of the observations were on AFDC. For the budget constraint, variables for the hourly wage rate ($W$), nonlabor income ($N$), and the AFDC guarantee and tax rate ($G$, $t$, and $r$) are needed.  To address the familiar problem of missing wages for nonworkers, a traditional selection model is estimated. Appendix Table \ref{tab:WageEqEst} reports estimates of this equation using OLS and a selection-bias adjustment.  The OLS coefficient estimates are almost identical to selection-adjusted estimates for most of the variables, but not all.  We will use the OLS estimates for our main analysis and then estimate the model with the selection-bias adjusted estimates as a sensitivity test.  For $N$, the weekly value of nontransfer nonlabor income reported in the survey is used.  AFDC guarantees and tax rates by year, state, and family size are taken from estimates by \cite{Ziliak2007}, who used administrative caseload data to estimate ``effective" guarantees and tax rates.  The effective guarantees and tax rates in the AFDC program differ from the nominal rates because the benefit formula has numerous exclusions and deductions which generate regions of zero tax rates and others with positive values but below the nominal rates because of earnings-related deductions. A long literature has used estimated effective guarantees and tax rates by regression methods, which are more accurate approximations to the parameters actually faced by recipients.\footnote{See the references in Ziliak for the long prior literature.}  The mean effective tax rate on earnings across years is approximately 0.41, considerably below the nominal rate of 1.0, and that on unearned income is approximately 0.30, also far below 1.0.\footnote{Both $G$ and $t$ have major cross-sectional variation, with the 1988 $G$ for a family of 3 ranging from \$100 per month to \$753 per month, and with the effective tax rate on earnings ranging from 0.12 to 0.66. The tax rate on unearned income also has a wide range, but it was invariably insignificant in the empirical analysis and hence is not represented in the estimations reported in the next section.} The analysis also controls for the guaranteed benefit in the Food Stamp program, which was available over this period to both participants and nonparticipants in the AFDC program.  The Food Stamp guarantee is set at the national level and hence varies only by family size and year, and consequently has relatively little variation in the sample used here. Those benefits are assumed to be equivalent to cash, as most of the literature suggests.

\subsection{Instruments} 
\hspace{5ex} We require instruments $Z_{i}$ that proxy fixed costs of participation that affect participation but not labor supply directly and which  meet the mean independence conditions in equations \eqref{eq:19} and \eqref{eq:20}.  For these instruments we use measures of administrative barriers to participation in the AFDC program which varied widely across the states. Students of the AFDC program in the 1970s and 1980s know that there is a sizable literature, appearing mostly in social work journals, documenting non-financial administrative barriers to program participation over the period (\cite{HanHol1971}; \cite{PilMasCor1979}; \cite{BrodLip1983}; \cite{Lipsky1984}; \cite{Lindseyetal1989}; and \cite{Kramer1990}). This literature showed that administrative barriers were politically driven at the gubernatorial and state legislature level and were aimed at keeping caseloads in the program down.  The program was regulated by the federal government, which required states to use benefit formulas, asset tests, and family composition rules set by federal agencies, so simply altering benefit levels or tax rates in the programs to reduce caseloads was difficult. Instead, states were able subjectively interpret the rules for what types of income to count, whether an able-bodied spouse or partner was present, which assets to count, and other factors affecting eligibility. Heavy paperwork requirements on applicants were imposed and states used failure to complete the paperwork properly as a reason for denying applications (``mechanisms to limit services...through imposing costs and inconvenience on clients" \cite{Lipsky1984}, p. 8). 

Measures of these administrative barriers are available because the federal government began auditing the states in the mid-1970s to determine whether they were making errors in assessing eligibility.  The auditing teams sent to the states selected random samples of applicant records and calculated a set of error rates for each state.  Very few types of error rates were initially calculated but the number and type of errors collected grew in the early 1980s.  While some of the data on these error rates were published, some were unpublished but exist in the internal files of the Department of Health and Human Services and were obtained for this project.\footnote{The rates which were published appear in annual issues of the publication Quarterly Public Assistance Statistics in the 1980s and 1990s.} For the time period covered by our SIPP data, the data provide information on seven measures of state AFDC administrative actions which are potential correlates of non-financial administrative barriers:  the percent of eligibility denials that were made in error, the error rate from improperly denying requests for hearings and appeals, the percent of cases dismissed for eligibility reasons other than the grant amount, the overall percent of applications denied, the percent of applications denied for procedural reasons (usually interpreted as not complying with paperwork), the percent of cases resulting in an incorrect overpayment or underpayment, and the percent of cases resulting in an underpayment.  There are also error rates and percents of actions related to income, assets, or employment, but these are directly or indirectly related to the applicant's labor supply and earnings level and hence are not used.

The means and distributional statistics of the seven administrative barrier variables are shown in Table \ref{tab:AdminBars}.\footnote{The administrative variables bounce around from year to year for each state because the federal government only took a random sample of records each year. To reduce noise, we compute the average of each barrier for each state over the 1988--1992 period.  The next section of the paper will report results using a change in their value in the early 1990s.}  While the means of one of the variables is less than 1 percent, others range from 2 percent to 24 percent.  The cross-state variation is also wide, with some states making underpayment errors in over 10 percent of cases, procedural denial rates of almost 35 percent, and overall denial rates of almost 50 percent.

There are obvious threats to the validity of any purely cross-sectional state-level government policy instrument like that used here. States differ in many demographic and economic characteristics which could be correlated with these error rates, either because both are correlated with some underlying labor-supply-related state characteristic or because there might be direct reverse causality running from labor supply levels to administrative barriers.  We conduct four exercises to provide evidence on instrument validity.  The first two are traditional balance tests and falsification tests.  The former will determine whether the instruments are correlated with the observable characteristics of the women in our SIPP sample which, though testing only correlations with observables, nevertheless is suggestive.  The second will determine whether effects of the instruments on labor supply are the same in samples of women not eligible for the program, which will tell us if the instruments are correlated with general labor supply levels of the populations in the states.  The third and fourth exercises more fundamentally change the source of instrument variation.  In one, we test the theoretical hypothesis that the administrative barriers are exogenously politically driven by examining their correlation with the political party in power at the state level and, going beyond that, we use close election instruments to isolate near-exogeneity of political control.  In the other, we use federal legislation in 1989 that altered the federal monitoring process and which we find to have differential effects across states which permits a traditional difference-in-difference analysis.  We postpone the falsification tests and these latter two exercises until the next section of the paper because they support the simpler cross-sectional results reported in this section.

\subsubsection{First-Stage Estimates, Balance Tests, and Instrument Strength}

To generate first-stage estimates of the AFDC participation propensity score, we match the state of residence of each observation in our SIPP data to the state administrative barrier variables and estimate probit models for the probability of AFDC participation as a function of each of those variables separately.  As required by the theory (see equation \eqref{eq:4}), the four budget constraint variables are included in the equation ($W$, $W(1-t)$, $G$, and $N$), and we also include nine conventional socioeconomic and area characteristics to arrive at the $X_{i}$ vector denoted in equation \eqref{eq:25}.\footnote{These are age, black, family size, number of children less than 6, the state unemployment rate, the Food Stamp guarantee, and three regional dummies.}  The first column of Table \ref{tab:ZonAFDCPart} shows that five of the seven variables have negative effects on AFDC participation, consistent with expectations, and two have positive but small coefficients.  However, the standard errors are high.  But much of the reason is that the administrative barriers are highly correlated with each other, as shown by the fourth column which reports the R-squared from a regression of each of the barrier variables on the other six.  While one is quite low (0.10), most of the others are sizable and as high as 0.81.  Thus it appears that states which have high error rates on one measure tend to have have error rates on the others.  

This suggests that a better approach would be to treat the barrier variables as noisy measures of an underlying index and, with this approach, the last two rows show the results when using two averages of the seven barrier variables in the model.  One is the textbook inverse variance weighted average which is the lowest variance estimate of a true single variable in the presence of measures with independent mean-zero measurement error, and the other is a simple unweighted average, for purposes of a sensitivity test.  The impacts of these index variables on AFDC participation are much stronger than those of the individual barrier variables.  We will consequently use these indexes henceforth, but will also report MTE curve estimates for the individual barrier variables as well.

Estimates of the impacts of the individual barrier variables in the first stage equation are not really correct, in any case, because the impact of each should be conditional on the other six (\cite{HeckVyt2005}, p. 700).  In general, as noted by \cite{MogTorWal2021}, different instruments may operate at different margins of the outcome variable (hours of work in our case).  Following the suggestion of those authors, the last two columns in Table \ref{tab:ZonAFDCPart} shows the effect on AFDC participation of each of the seven barrier variables conditional on the other six.  These results are very weak, with small coefficients and high standard errors.  This is a result of the same collinearity problem just noted.  This suggests again that all the barrier variables are proxying approximately the same state behavior, and, indeed, our results in the MTE curve estimation reported below will show similar MTE curves for all the barrier variables individually, consistent with the assumption that they are all operating at about the same margin.

Table \ref{tab:FSPBCInts} tests whether the administrative barrier indices significantly interact with the four budget constraint variables.  The interactions are added sequentially. Three are statistically significant but, conditional on those three, gross wages have no additional explanatory power.  The signs on the interactions imply that those with higher levels of three of the budget constraint variables result in stronger negative effects of the barrier indices on program participation.  There is no theoretical prediction for these signs, because they reflect the density of utility gains in the region of the data where the instruments are operating, and the resulting fractions of women moved over the participation threshold for different levels of the instruments, and this is purely a function of the data.

For balance tests, we examine the balance between our administrative barrier indices and the 13 elements of our $X_{i}$ vector. Most correlation coefficients between our indices and those elements are low but several are above 0.3 in absolute value and one as high as 0.5 (Appendix Table \ref{tab:LogZCorrMat}).  To test the sensitivity of our MTE estimates to imbalance, we construct, and then condition on, propensity scores of $Z$ as a function of $X$ which improve balance.  With $Z$ a continuous variable, we apply the method suggested by \cite{HirImb2004} designed for models with a continuous treatment variable.\footnote{The method is designed to ensure unconfoundedness in an outcome equation, but here we apply the method to our first-stage equation.}  Following their approach, we construct a propensity score by estimating an equation for the density of $Z$ as a nonlinear function of our $X_{i}$ vector and then look for specifications of that function which result in good balance between our instruments $Z$ and our covariate vector $X$ after conditioning on those estimated propensity scores. Balance is gauged by how many covariate means are significantly different across percentile intervals of $Z$ after conditioning on the score.  Table \ref{tab:OPHIBalSum} shows four different propensity score specifications that improve balance considerably.\footnote{Appendix Table \ref{tab:OPHIBal} details the changes in individual covariate balance before and after adjusting for the propensity score of $Z$.}  We will estimate our MTE curves in the next section after conditioning on these four propensity scores in both the first and second stages of the model to test for the sensitivity to covariate imbalance.

For instrument strength, we note that the rules-of-thumb F-statistics employed in much current IV work are not directly relevant for our work.  Those rules of thumb gauge instrument strength for a single LATE estimate in the outcome equation, not for an MTE curve.  For an MTE curve, the relevant question is not whether the weighted average of MTEs that constitutes a LATE (\cite{HeckVyt2005}) has strength, but how much strength the instruments have at each point on an MTE curve.  There are no results in the existing weak IV literature for strength calculations at each point on a continuous MTE curve, so we instead apply the results from the single LATE estimate literature to different segments of the propensity score distribution.  We can thereby gauge the strength and weakness of our instrument separately in different ranges of the curve.

\begin{figure}[ht]	
	\caption{Histogram of Predicted Participation Rates}\label{fig:PredPartRates}
	\centering
	\begin{minipage}{\linewidth}
		\includegraphics[scale=1.0]{"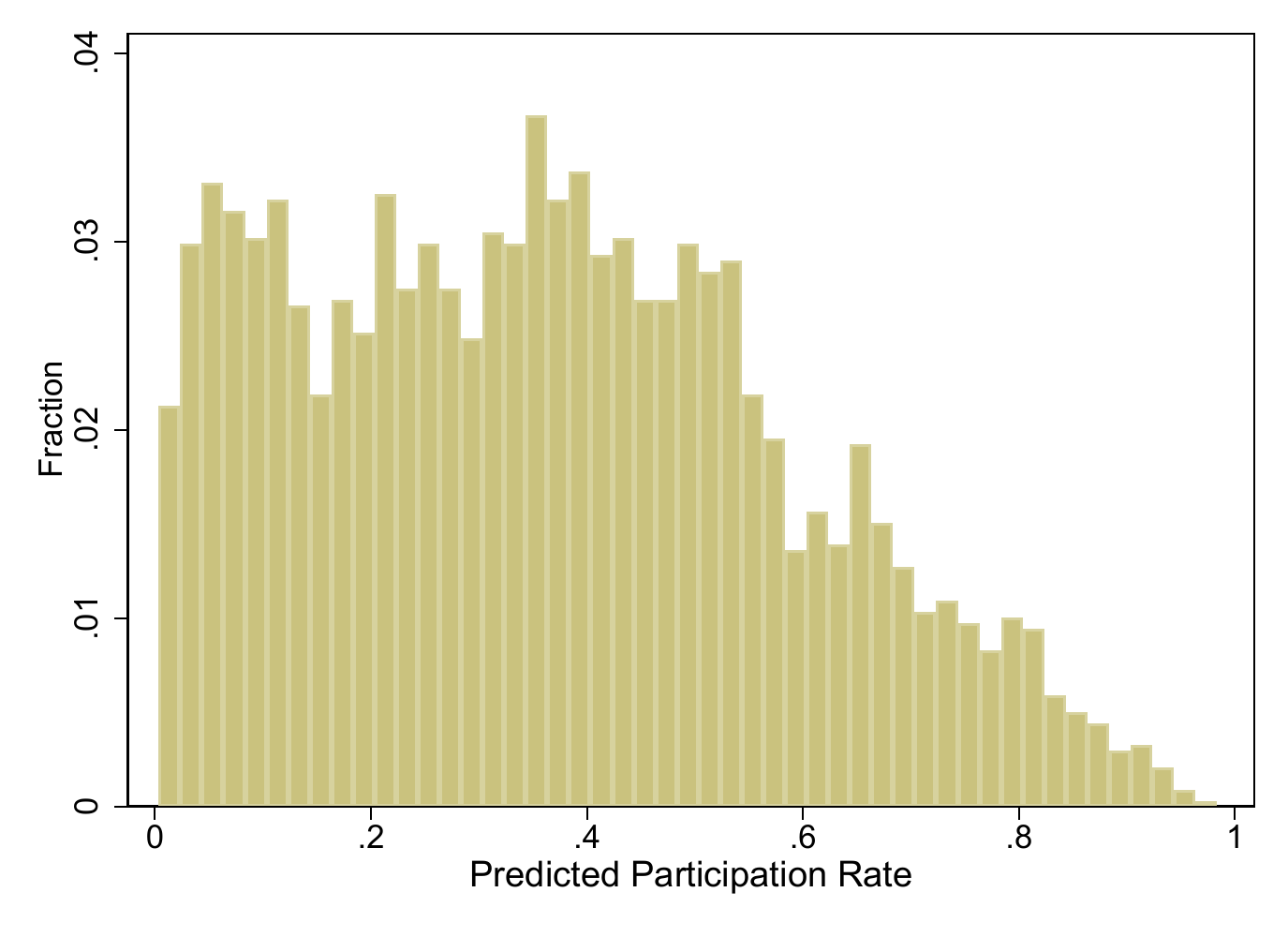"}
		\rule{\linewidth}{0in}
		\footnotesize
		\emph{Notes:} This figure plots the predicted participation probabilities from the AFDC probit using the inverse variance weighted average of the logged administrative barrier instruments and no interactions.  
	\end{minipage}
\end{figure}

Figure \ref{fig:PredPartRates} shows the distribution of estimated propensity scores in our sample.  The density fluctuates around approximately the same level between 0 and about 0.55, then declines at a steady pace almost up to 1.0.  We divide the propensity score into terciles and quartiles and calculate separate F-statistics for the instruments within each range.  The results are shown in Table \ref{tab:pFSum} for our two barrier indices and for the first four specifications in Table \ref{tab:FSPBCInts} with significant effects of the instruments.  The tercile results show that the instruments are weak in the bottom and top tercile, while the quartile results show weakness in the instrument in the first, third, and fourth quartiles.  The instruments with interactions with the budget constraint variables are strongest in the approximate range 0.25 to 0.66.\footnote{It should not be surprising that instruments are strongest in the middle of the distribution and weakest in the tails because the common S-shaped cdf curve for most distributions generates the steepest slope in the middle and the flattest slope in the tails.}

While the F-statistic rule-of-10 for strong instruments is commonly used (arguably overused), and usually applied independent of the particular application and distribution of the data, the simulations of \cite{StaSto1997} and \cite{StoYog2005}, upon which the rule is based, are motivated by the case of many weak instruments (as in the work of \cite{AngKru1991} which prompted the literature) rather than the just-identified, single instrument case used here.  In fact, in the just-identified case, \cite{KeaNea2021} and \cite{AngKol2021} show that, unless the bias of OLS is extraordinarily high, median bias and undercoverage of the 2SLS estimator is quite minor for almost any F-statistic. This occurs because, as the first-stage F-statistic falls, the confidence interval of the estimator widens, reducing undercoverage. On the other hand, while these results apply to bias and coverage, \cite{KeaNea2021} and \cite{Leeetal2021} show that power of the 2SLS estimator is lower than that conveyed by conventional t-statistic rules even in the just-identified case. Nevertheless, given the lack of work on the MTE case we are concerned with, we will proceed in our empirical work to restrict our attention to the MTE curve in the (0.25, 0.66) range and, in the next section, we will test the sensitivity of our estimates to the use of alternative instruments.

\subsection{Main Results}
\hspace{5ex}Estimation of equation \eqref{eq:24} using the fitted values of the participation probabilities for $F$ yields estimates of $\beta$, $\lambda$, and the parameters of the $g$ function.  The $g$ function is estimated with conventional cubic splines, hence $g(F)=g_{0}+\sum\limits_{j=1}^{J}g_{j}\max(0,F-\pi _{j})^{3}$, where the $\pi _{j}$ are $J$ preset spline knots.  For a given $J$, the knots will are chosen to be regularly spaced within  the (0.25, 0.66) range. The estimation will start with $J=3$ and then increase the number until a fit measure is maximized. Fit will be assessed with a generalized cross-validation statistic (GCV).  Given the well-known tendency of polynomials to reach implausible values in the tails of the function and beyond the range of the data, natural splines are typically used, which constrain the function to be linear before the first knot and beyond the last knot (\cite{Hastieetal2009}).  Imposing linearity on the function in those two intervals requires modifying the spline functions to accommodate this; the exact spline functions for a five-knot spline are shown in \ref{app:spline}.\footnote{Consistency of sieve methods is discussed by \cite{Chen2007}.}

\begin{figure}[ht]	
	\caption{Marginal Labor Supply Curves for Different Natural Cubic Splines}\label{fig:MTEKnotCompInvVarWtg}
	\hspace*{-2.0cm}
	\centering
	\begin{minipage}{\linewidth}
		\includegraphics[scale=0.80]{"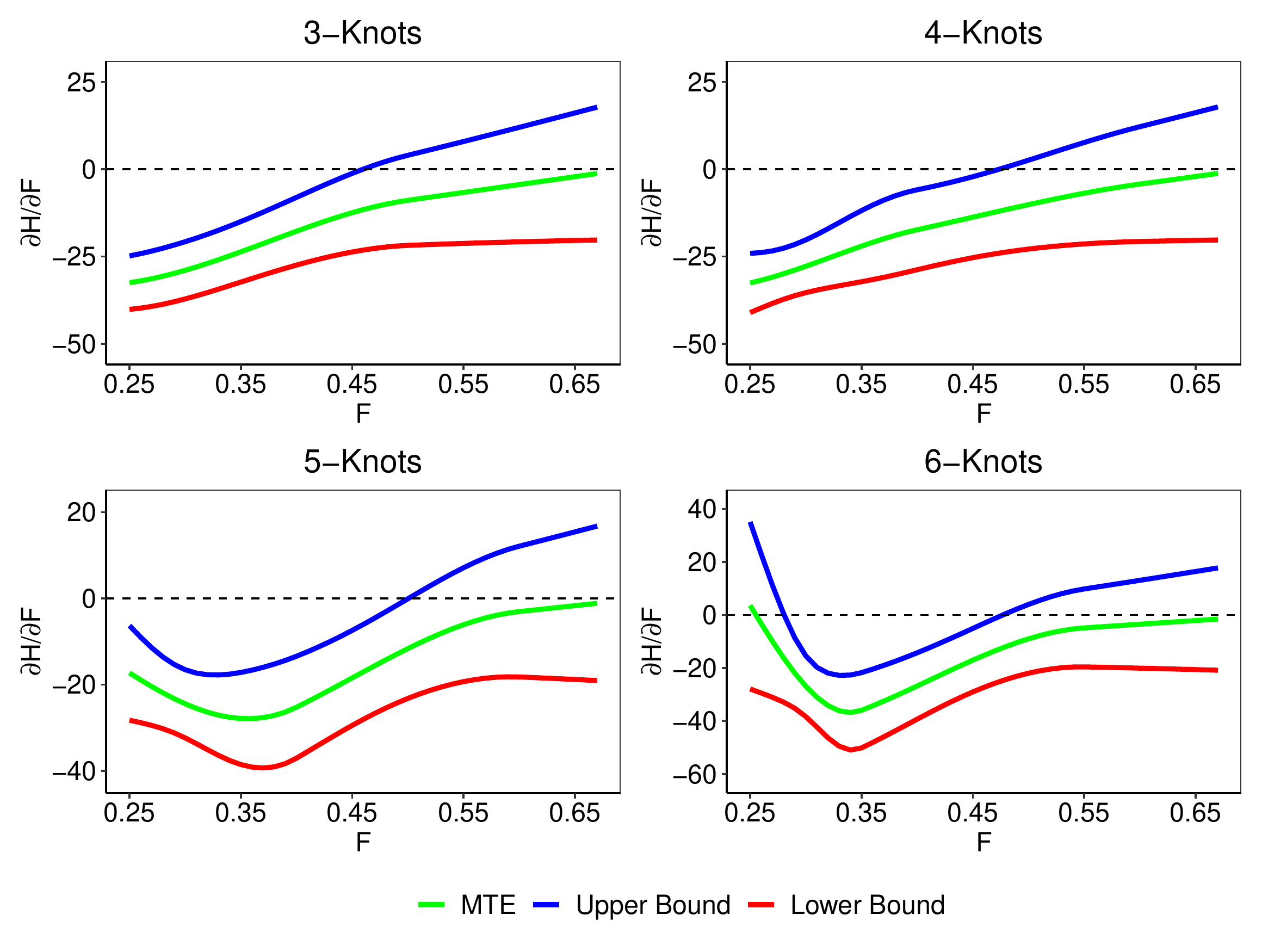"}
		\rule{\linewidth}{0in}
		\footnotesize
		\emph{Notes:} This figure plots the marginal treatment effect curves using different cubic spline specifications. All specifications use a first stage probit model with the inverse variance weighted log of the AFDC administrative barriers and interactions with $N$, $G$, and $W(1-t)$.  Upper and lower bounds are generated from a block bootstrap with replacement at the state level using 500 samples.
	\end{minipage}
\end{figure}

Figure \ref{fig:MTEKnotCompInvVarWtg} shows the estimated MTE curves in the (0.25, 0.66) range with 95 percent confidence intervals for a specification using the inverse variance weighted barrier index with three budget constraint interactions (hence column (4) in Table \ref{tab:FSPBCInts}), and for three-to-six knots.\footnote{The MTE function is, as noted previously, just the derivative of the hours equation w.r.t the participation rate. The confidence intervals are constructed using block bootstrap methods which allow for state-specific clustering. All MTE curves are evaluated at the means of the other variables in the equation.}  As shown in Table \ref{tab:GCVSum}, the GCV hits its minimum at 5 knots for this specification but its GCV and that for 6 knots are almost identical, so both are shown. For these knot specifications, the marginal responses are non-monotonic and U-shaped, starting off at $F=0.25$ significantly different from 0 in the 5-knot case but insignficantly different in the 6-knot case, but then growing in (negative) size as participation increases.  The marginal response peaks at a participation probability in the (0.30, 0.40) range, depending on the specification, of about 0.35, when it reaches approximately -30 to -40 hours per week. It then declines, becoming insignificantly different from 0 at approximately $F=0.47$.  The point estimate approaches zero as participation rises further but remains insignificantly different from 0 for all higher participation levels.  Appendix Figure \ref{fig:MTEKnotCompSimpAvg} shows the same figure for the simple average barrier index, with the MTE curves almost identical in shape.\footnote{Appendix Figure \ref{fig:MTEHeck} shows the 5-knot and 6-knot MTE curves using the Heckman selection-bias-adjusted hourly wage rate.  The shapes and confidence intervals are quite close to those in Figure \ref{fig:MTEKnotCompInvVarWtg} using OLS-predicted wages.}

Table \ref{tab:HoursEqSplineInt} shows the full set of parameter estimates for three versions of the hours equation for the 5-knot specification.  The natural spline coefficients are not easily interpretable and instead are only shown graphically in Figure \ref{fig:MTEKnotCompInvVarWtg}.  Column (1) has only the budget constraint variables in the $\lambda$ vector, which are not very strong predictors of hours, implying that we do not detect strong interactions of participation with those variables. The wage itself does have strong positive effects on hours, however, as indicated by its $\beta$ coefficient. Column (2) tests a set of additional interactions of the participation probability with the budget constraint variables, but no effects are found there.  We tested additional $X$ variables in the $\lambda$ vector and column (3) shows the results when Age and Black are added, which were marginally significant in various specifications but insignificant in the one shown, but always improve the GCV measure. Other variables in the $X$ vector also did not enter significantly. The spline coefficients in column (3) are those used in Figure \ref{fig:MTEKnotCompInvVarWtg}. But we show in Appendix Figure \ref{fig:MTEDiffH5k} the estimated MTE curves from the specifications in Columns (1) and (2), which are very close to those in Figure \ref{fig:MTEKnotCompInvVarWtg}, implying that these specification issues do not affect our general MTE results.

\begin{figure}[ht]	
	\caption{Marginal Labor Supply Curves for Six-Knot Model with Different Interactions}\label{fig:MTEIntComp6K}
	\hspace*{-2.0cm}
	\centering
	\begin{minipage}{\linewidth}
		\includegraphics[scale=0.80]{"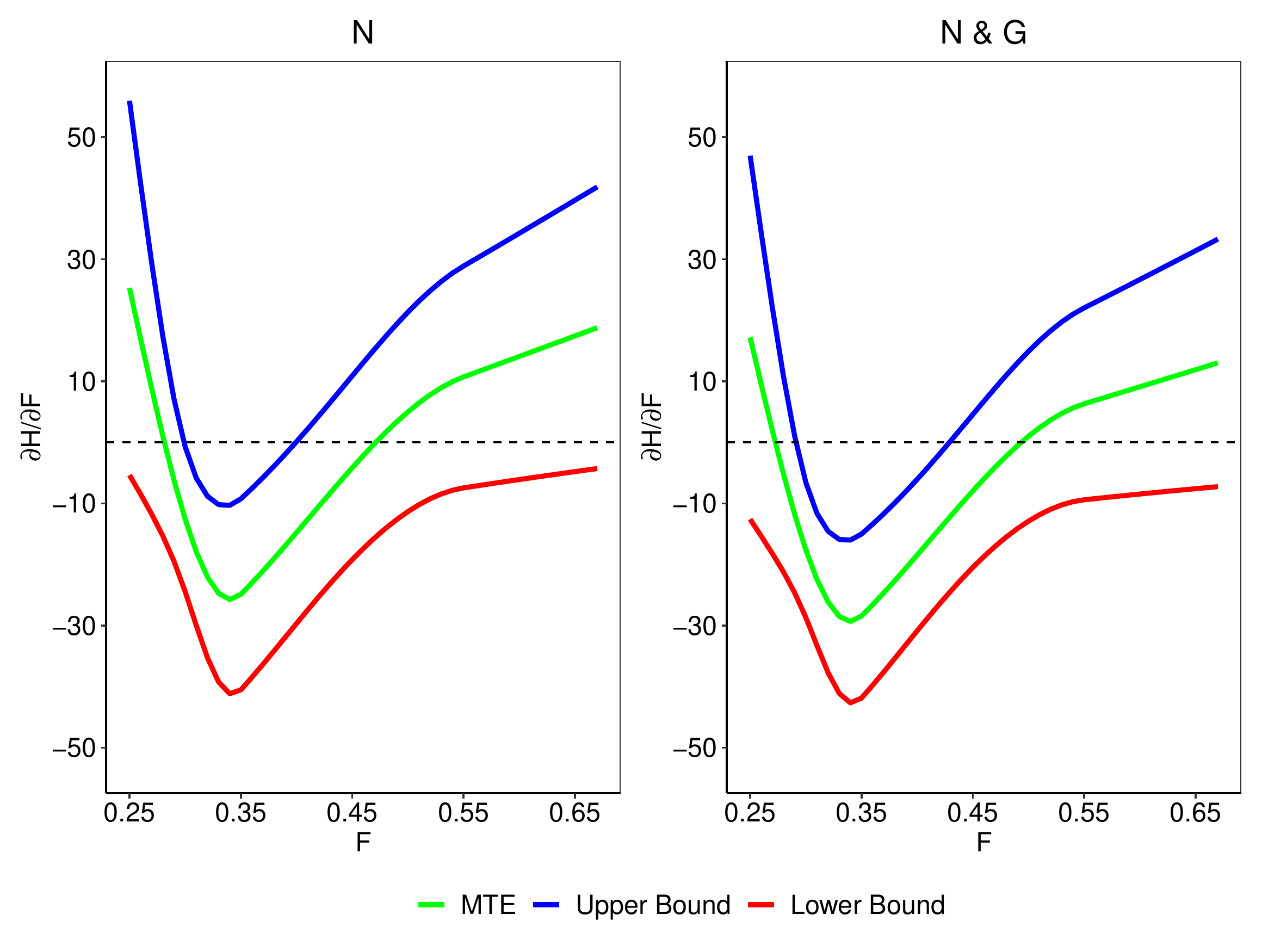"}
		\rule{\linewidth}{0in}
		\footnotesize
		\emph{Notes:} This figure plots the marginal treatment effect curves using a 6-knot cubic spline specification and a first stage probit model with the inverse variance weighted log of the AFDC administrative barriers and different interactions noted above the graph. Upper and lower bounds are generated from a block bootstrap with replacement at the state level using 500 samples. 
	\end{minipage}
\end{figure}

Figure \ref{fig:MTEIntComp6K} shows the MTE curves for the other two budget constraint interaction specifications in the first-stage (specifically, columns (2) and (3) in Table \ref{tab:FSPBCInts}), for the 6-knot specification where their GCV is minimized.  The general shape of the curves is the same as Figure \ref{fig:MTEKnotCompInvVarWtg}, although the points at which the MTE is significantly different from zero shift slightly because of shifts in the 95 percent confidence interval.  The peak work disincentive point estimate is about -25 to -30 hours per week (depending on the specification), a bit smaller than some of the curves in Figure \ref{fig:MTEKnotCompInvVarWtg} but quite close. 

\begin{figure}[ht]	
	\caption{Marginal Labor Supply Curves Controlling for the GPS}\label{fig:MTEGPSLogZ5K}
	\hspace*{-2.5cm}
	\centering
	\begin{minipage}{\linewidth}
		\includegraphics[scale=0.85]{"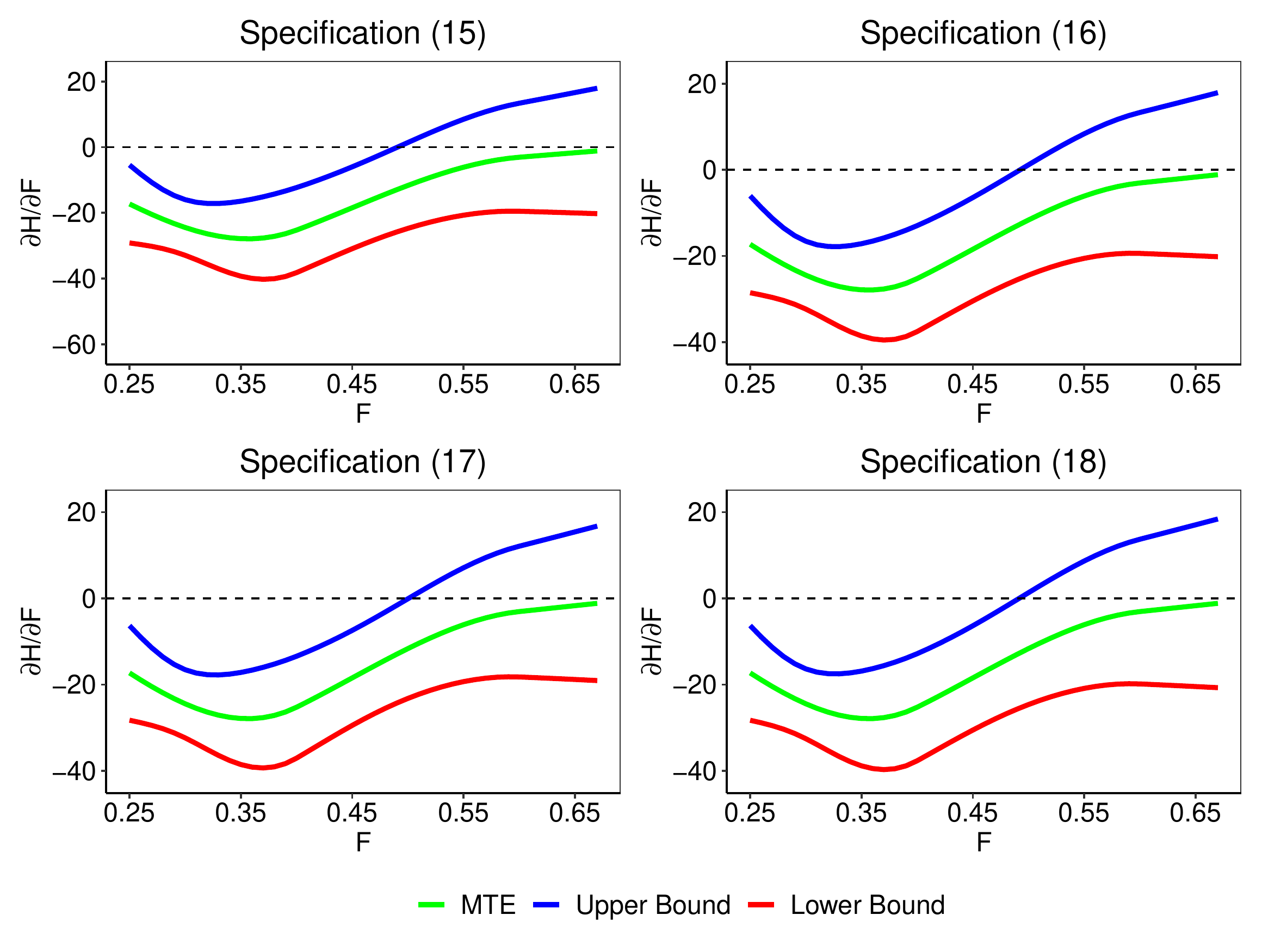"}
		\rule{\linewidth}{0in}
		\footnotesize
		\emph{Notes:} This figure plots the marginal treatment effect curves using a 5-knot cubic spline specification and a first stage probit model with different administrative barrier instruments in logs and interactions with $N$, $G$, and $W(1-t)$. The first stage probit and the second stage hours equation control for the GPS generated from the specifications presented in Table \ref{tab:OPHIBalSum}. Upper and lower bounds are generated from a block bootstrap with replacement at the state level using 500 samples.  
	\end{minipage}
\end{figure}

Figure \ref{fig:MTEGPSLogZ5K} conducts sensitivity tests to covariate balance by showing the effect on the 5-knot MTE curve in Figure \ref{fig:MTEKnotCompInvVarWtg} when the four different GPS variables are included in the first- and second-stages of the model.  All four specifications of the GPS in Table \ref{tab:OPHIBalSum} are shown (although only for the base specification using the inverse variance weighted barrier index). The results are virtually unchanged from those without conditioning on the score, showing that improvements in covariate balance, at least in the ranges implied by Table \ref{tab:OPHIBalSum}, have no effect on our estimated MTE curves.
 
Estimation of a homogeneous effects model, equivalent to specifying the $g$ function as a constant, yields a point estimate of -25 hours per week (s.e.=6.6).  As is well known, linear IV assigns weights to the different MTEs at different points in the propensity score distribution (\cite{HeckVyt1999,HeckVyt2001,HeckVyt2005} and \cite{AngGradImb2000}). In this application, the weights are concentrated around 0.35.  Linear IV would therefore give a wildly distorted picture of how marginal responses vary and would completely miss the U-shaped response function which actually occurs.  And, most important, it would miss the main implication of the results so far---that the labor supply effects of AFDC expansions at the margin are often small and insignificantly different from zero, but in some regions of program expansion they can be very large, with important policy implications.

\begin{figure}[ht]	
	\caption{Marginal Labor Supply Curves for Different Administrative Barrier Instruments}\label{fig:MTEDiffLogZ5k}
	\hspace*{-2.5cm}
	\centering
	\begin{minipage}{\linewidth}
		\includegraphics[scale=0.85]{"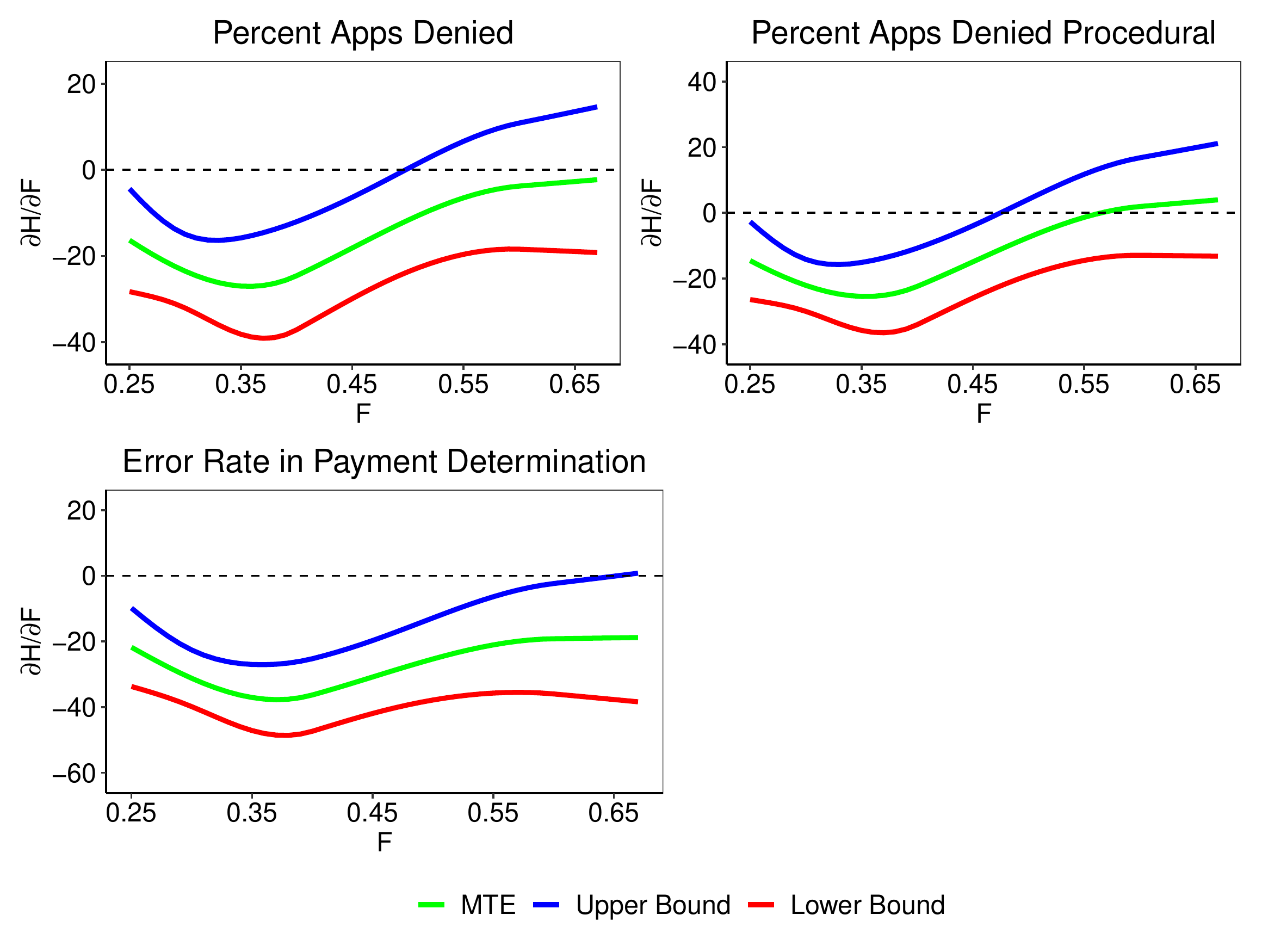"}
		\rule{\linewidth}{0in}
		\footnotesize
		\emph{Notes:} This figure plots the marginal treatment effect curves using a 5-knot cubic spline specification and a first stage probit model with different administrative barrier instruments in logs and interactions with $N$, $G$, and $W(1-t)$.  Upper and lower bounds are generated from a block bootstrap with replacement at the state level using 500 samples.
	\end{minipage}
\end{figure}

Figure \ref{fig:MTEDiffLogZ5k} shows the MTE estimates for the three individual barrier instruments that have the highest F-statistics in the (0.25, 0.66) range, using interactions with the three significant budget constraint variables and 5-knots.\footnote{Appendix Table \ref{tab:pFSumLogZs} shows the tercile and quartile F-statistics for each of the 7 individual barrier variables for this specification.} The estimated MTE curves are almost identical to those for the barrier index variables, supporting our hypothesis that all are picking up approximately the same behavior and operating at approximately the same margin of marginal labor supply disincentives in the population.

We end this section on our main results by an attempt to gain some insight into the mechanics behind the U-shaped pattern of responses we have found.  We conduct two exercises.  The first examines where in the distribution of hours the responses come from over different participation ranges---in particular, by examining how individuals reduce hours from 40 per week or 20 per week or to lower levels, including non-work.  That movements between full-time work, part-time work, and non-work may be important is demonstrated in Table \ref{tab:ShareHbyP}, which shows the distribution of welfare participants and non-participants across the hours categories.  What is striking about the table is that welfare participation is essentially equivalent to not working, with almost no participants working part-time and even fewer working full-time. Among non-recipients, the distribution is the opposite, with almost everyone working and over 80 percent working full-time.  While these distributions are not causal, they suggest that being off welfare is generally associated with working full-time and being on welfare is generally associated with not working, and that some of those who go onto welfare may reduce their hours by 40 per week.

\begin{figure}[ht]	
	\caption{Marginal Labor Supply Curves for Different Types of Workers}\label{fig:MTEHComp}
	\hspace*{-2.0cm}
	\centering
	\begin{minipage}{\linewidth}
		\includegraphics[scale=0.80]{"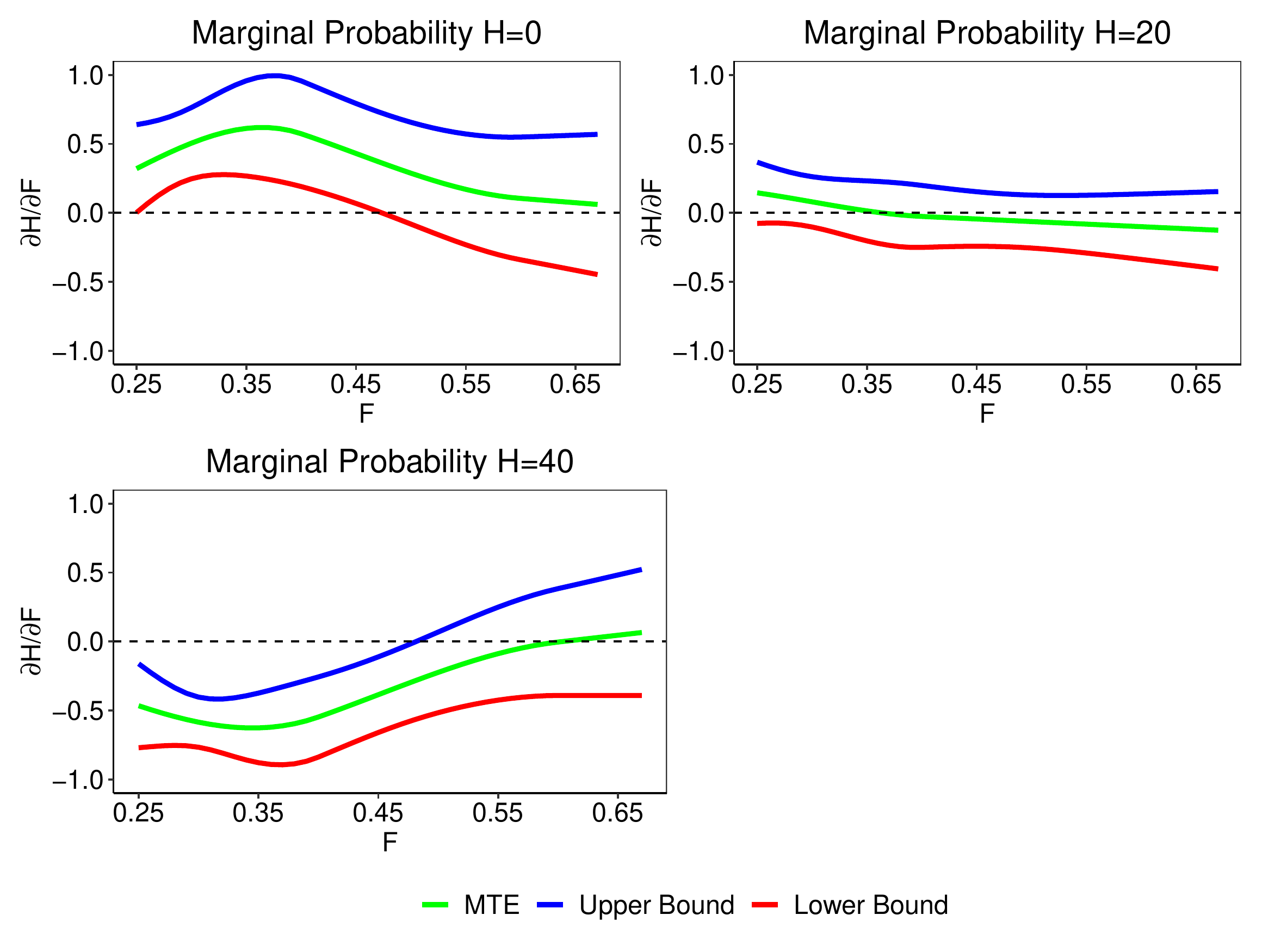"}
		\rule{\linewidth}{0in}
		\footnotesize
		\emph{Notes:} This figure plots the marginal treatment effect curves for none workers, part-time, and full-time workers using a 5-knot cubic spline specification and a first stage probit model with the inverse variance weighted log of the AFDC administrative barriers and interactions with $N$, $G$, and $W(1-t)$. Upper and lower bounds are generated from a block bootstrap with replacement at the state level using 500 samples.  
	\end{minipage}
\end{figure}

Evidence suggesting this is the case is shown in Figure \ref{fig:MTEHComp}, which shows the result of estimating the hours worked equation by successively replacing the dependent variable for $H$ with dummies for not working, working part-time, and working full-time.  The figure shows the MTEs from those regressions.  The leftmost panel shows that the probability of nonwork rises sharply as participation goes from 0.25 to 0.35, the same range where the MTE for average hours falls the most.  The middle panel shows that the MTE for part-time work actually starts off at a positive level (albeit small), implying an increase in part-time work that can only come from full-time workers reducing labor supply to the part-time level.  The part-time MTE becomes less positive as participation increases and eventually becomes zero or negative, implying that some part-timers move at that point to nonwork.  But the right panel shows that the MTE for full-time work is large and negative in the 0.25 to 0.35 participation rate range implying, when combined with the other panels, that a large part of the reduction in labor supply over that range is from full-time work to nonwork upon participation, which is where the prior figures show the maximum reduction occurs.  Eventually, however, after participation rises high enough, movements out of full-time work fall to zero.  Thus the decline in the labor supply reductions in average hours when participation rates rise sufficiently reflects a decline in movements out of full-time work.

Further evidence that it is the high-hours-worked individuals who participate ``early" (i.e., when administrative barriers and fixed costs are high and hence participation is low) who are responsible for the large marginal effects in the lower ranges of the participation rate distribution is shown in Table \ref{tab:VarMeansbyFZ}, which displays a few labor-supply related variables by quintile of the fitted propensity score distribution within the (0.25, 0.66) range.   Those who are on the margin at low participation probabilities have higher wage rates, are less likely to be black, are older, and have fewer young children, all of which are correlated with higher levels of work.  Nonlabor income is higher for the early participants as well, which is typically correlated with lower levels of labor supply but, for discrete moves from full time work to nonwork, this means that those individuals also have a larger income cushion if they do not work.  Those who are on the margin at higher participation rates have lower wages, are more likely to be black, are younger, and have more children, all of which are correlated with lower levels of work and hence lower marginal effects of labor supply upon participation.\footnote{It may be worth noting that these patterns are not implied by the theoretical model and hence do not have to come out this way. The model shows that who participates early and who participates late is entirely a matter of relative preferences for leisure and consumption goods, and those relative preferences can vary arbitrarily in the preference distribution. Consequently, the pattern in Table \ref{tab:VarMeansbyFZ} is a substantive finding that helps interpret the U-shaped MTE pattern we have found.} \footnote{This method of examining heterogeneous response from differences in observable characteristics at different percentile points is closely related to the method recently suggested by \cite{CherFernLuo2019}.}

\section{Sensitivity Tests to Instrument Validity}
We conduct three sensitivity tests to the validity of the instruments:  falsification tests, a difference-in-difference exercise, and a test using a political, close election regression discontinuity design.

\subsection{Falsification Tests}

\begin{figure}[ht]	
	\caption{Marginal Labor Supply Curves for Different Samples}\label{fig:MTEDiffSamp}
	\hspace*{-2.0cm}
	\centering
	\begin{minipage}{\linewidth}
		\includegraphics[scale=0.80]{"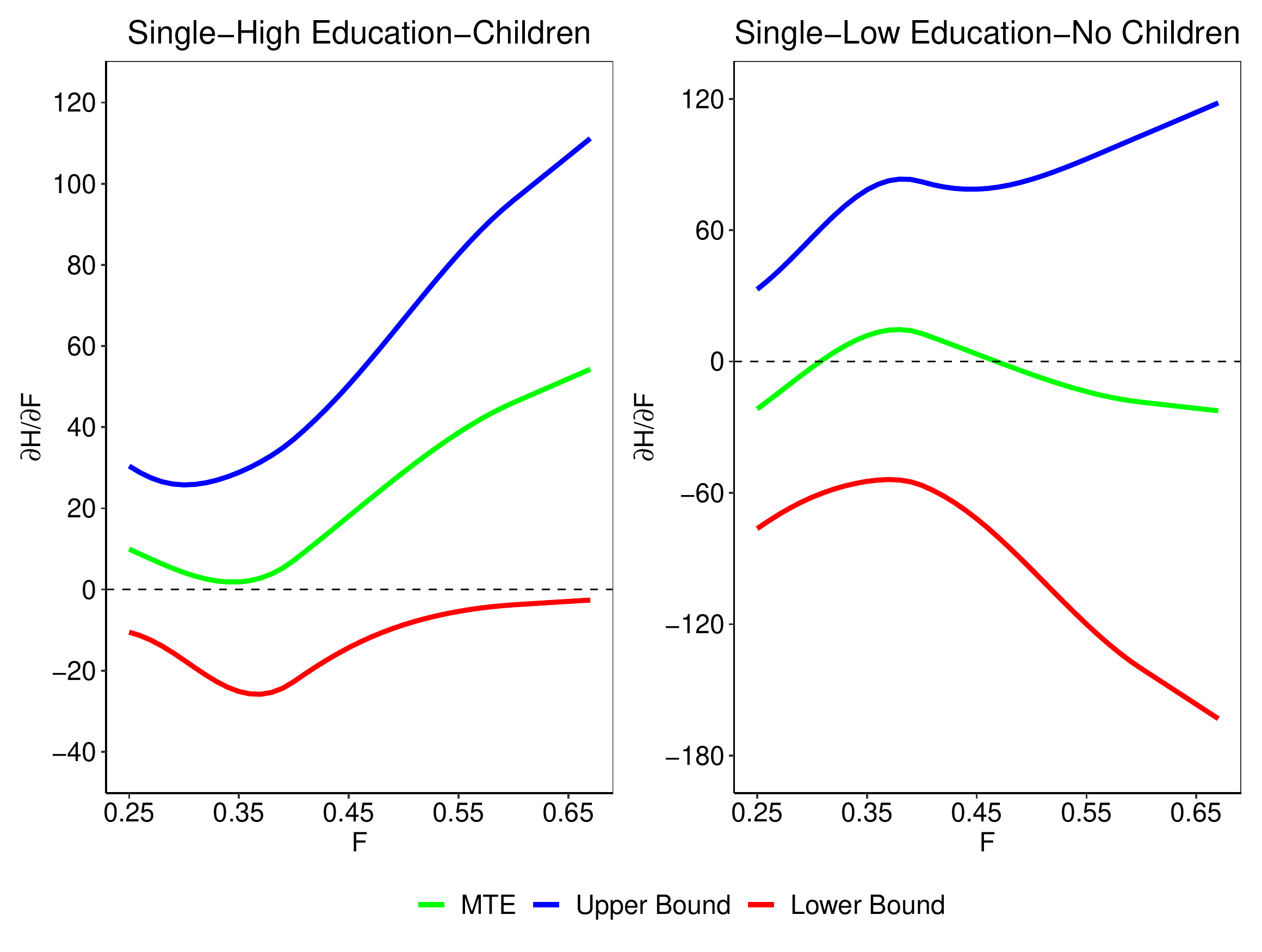"}
		\rule{\linewidth}{0in}
		\footnotesize
		\emph{Notes:} This figure plots the marginal treatment effect curves for different samples of women using a 5-knot cubic spline specification. AFDC participation probabilities use the parameters from our preferred probit specification using the inverse variance weighted log of the AFDC administrative barriers and interactions with $N$, $G$, and $W(1-t)$ and the data values for the women in our alternate samples. Upper and lower bounds are generated from a block bootstrap with replacement at the state level using 500 samples.  
	\end{minipage}
\end{figure}

Our sample used for the main results consists of low-education, low-asset single mothers.  We estimate our baseline model on two alternative samples: high education single women with children and low education single women without children.  These groups are essentially ineligible for AFDC and hence the AFDC administrative barrier variables should have no effect in their labor supply.  To implement this test in our LIV model, we use the estimated parameters from the AFDC participation probit estimated on single mothers, but predict the propensity score using the covariates for the women in each alternative sample.  With the control variables in the $H$ equation also replaced by their values for the women in each alternative sample, the estimated MTE from that equation will reflect the variation in the administrative barrier indices.  The estimated MTE curves and associated 95 percent confidence intervals from this exercise are shown in Figure \ref{fig:MTEDiffSamp}.  In both alternative samples, the implied effects have wide confidence intervals and are insignificantly different over the entire range of propensity scores.

\subsection{Difference in Difference Test}
Congress passed new legislation in 1989, the Omnibus Budget Reconciliation Act, which modified the quality control inspection program that the federal government used to assess state error rates (\cite{USGreenBook1994}, Section 10). The legislation was motivated by a concern that states were continuing to make errors in their program eligibility assessments and tightened up the monitoring system imposed on the states.  The full implementation of the Act started in late 1991 and was completed in 1992.  We use this legislation in a difference-in-difference exercise which examines whether error rates in the states changed significantly in 1992 compared to previous levels, and whether it did so differentially across states.  We then use that cross-state differential change in error rates as the instrument for estimating our MTE curve.

While this exercise serves as a worthwhile test, it is inferior to the variation used in our main analysis in two respects.  First, it was national legislation that was supposed to apply uniformly to all states, so we cannot determine the reason that different states reacted differently to the legislation.  This means that the test is something of a black box.  Second, we only have one year of data for the ``post" impact of the legislation---1992 is our last observation year---which reduces the power of the test.

Table \ref{tab:LogZTSMicroReg} shows the results of several regressions examining the impact of the 1992 discontinuity.  The first column shows, for illustration, the coefficients on a linear time trend variable and a 1992 dummy variable in a regression of our state- and year-specific administrative barrier index on those variables and the other control variables in the first-stage AFDC participation probit.  The 1992 dummy is positive and statistically significant at conventional levels, indicating that administrative barriers showed a positive deviation from trend in 1992.  We interpret this result as likely reflecting an increased detection of errors in the states.  The second column reports the coefficients on a time trend and a 1992 indicator in our AFDC participation probit, showing a statistically significant decline in AFDC participation in 1992 relative to trend, which we interpret as a result of the increase in administrative barriers.  The MTE curve that results from using this first-stage equation, using the 1992 indicator as the instrument, is shown in Appendix Figure \ref{fig:MTETimeSeriesAlt} and is close to that in our main results.

\begin{figure}[ht]	
	\caption{Marginal Labor Supply Curves Using 1992 Law Change Instrument}\label{fig:MTETimeSeries}
	\hspace*{-2.0cm}
	\centering
	\begin{minipage}{\linewidth}
		\includegraphics[scale=0.80]{"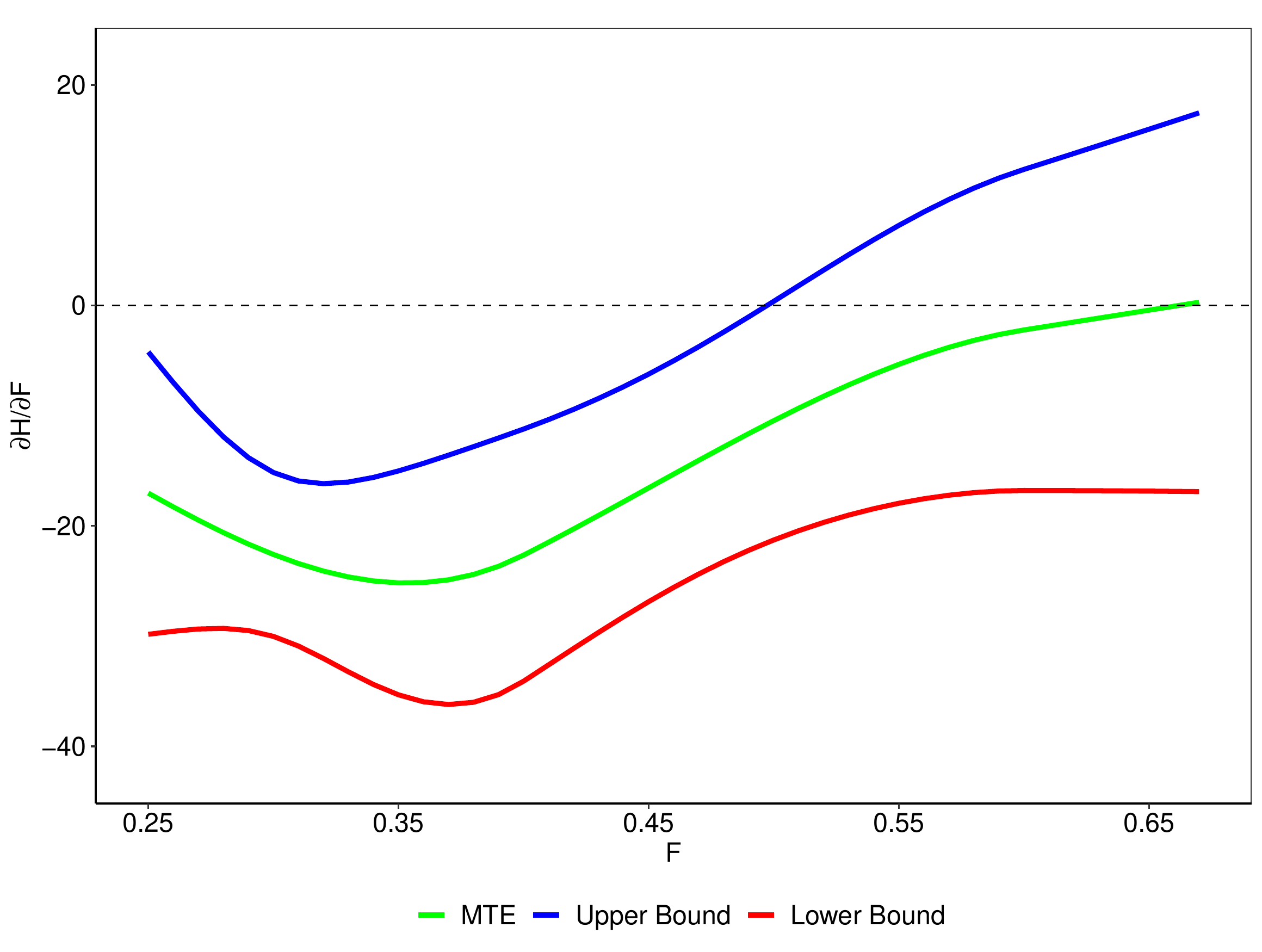"}
		\rule{\linewidth}{0in}
		\footnotesize
		\emph{Notes:} This figure plots the marginal treatment effect curves using a 5-knot cubic spline specification and the first stage probit model using a predicted value for the inverse variance weighted log of the AFDC administrative barrier instruments in column (3) of Table \ref{tab:LogZTSMicroReg}. Upper and lower bounds are generated from a block bootstrap with replacement at the state level using 500 samples.   
	\end{minipage}
\end{figure}

But this approach uses the pure time-series variation in AFDC participation for identification and does not use cross-state variation in the change in administrative barriers across states.  We use that variation by first estimating a regression for the log Z in each state separately on a time trend, and then predict the trending log Z for each state separately over the 1988--1992 period. Adding those predicted state-specific time trends in log Z to the AFDC participation probit, but also adding the 1992 residual from those state-specific regressions, we can estimate the impact on AFDC participation of state-specific 1992 deviations in the administrative barriers.  The last column of Table \ref{tab:LogZTSMicroReg} shows that the impact of the 1992 residual is negative but of low significance, no doubt partly because of the loss of power from small sample sizes by state and hence noisy estimates of the trend and 1992 deviation.  Nevertheless, the MTE curves obtained when using that state-specific deviation in the hours equation are shown in Figure \ref{fig:MTETimeSeries}.\footnote{The state-specific linear log Z trend is included in the hours equation to ensure identification solely from the 1992 deviation.} The estimated MTE curve is very similar to that in our main results and, in fact, has 95 percent significance in approximately the same range.

\subsection{Close Election RD Test}
As noted in our initial discussion of the administrative barrier instruments, their cross-state variation is argued by many researchers to be a result of political differences across the states.  But, as is widely recognized, political differences themselves may not be valid instruments because they are likely correlated with state demographics and therefore possibly with the labor market participation levels of low income families. We draw upon the literature on regression discontinuity designs in political economy research which use close elections as a plausibly exogenous source of political party governance (\cite{LeeMorBut2004}, \cite{Lee2008}, and the large subsequent literature).  The argument in this approach  is that states where a party is elected only narrowly is close in unobserved ways to states where parties lose narrowly, and therefore a comparison of the impact of which party is elected in a close election has a better chance of exogeneity than merely political party control itself, which could easily be correlated with state demographics.

We collect data on the party affiliation of the governor of each state in our data in our covered years, and we determine whether that governor was a Democrat elected in a close election, which we define alternatively as having been elected with either 50--55 percent of the vote or 50--60 percent, as a sensitivity test.  We control for the Democratic share of the vote as the running variable.  

We also gather information on the political makeup of the state legislature, which should affect the ability of governors to enact policies of their liking.  We collect data on whether the legislature is entirely Republican or whether it is split, with one chamber controlled by Democrats and one controlled by Republicans (a ``split" legislature).  We will test whether the impact of a Democrat governor who has been elected in a close election varies with these legislative party control variables.

The first two columns of Table \ref{tab:FSPPolVar} show the results of OLS regressions of our administrative barrier index on various political variables (plus the usual first stage control variables), including the close election Democratic gubernatorial variable.  The close election variable has a negative impact on the level of administrative barriers in the state, but is insignificant. However, we find that if that close election takes place when the legislature is controlled by the Republican party there is a large significant positive impact on the level of administrative barriers in the state for the closer election.  This could be because governors which are elected with bare majorities have weak political power relative to an established legislature controlled by the other party.  Columns (3) and (4) show the impact of the close election variable on AFDC participation in our data and show that, while the uninteracted close election variables have insignificant coefficients in those equations, the interaction terms are large in magnitude and negative, implying that Republican controlled legislatures result in reduced AFDC participation when operating with weak Democrat governors.

\begin{figure}[ht]	
	\caption{Marginal Labor Supply Curves Controlling for Political Variables}\label{fig:MTEPolVars}
	\hspace*{-2.0cm}
	\centering
	\begin{minipage}{\linewidth}
		\includegraphics[scale=0.80]{"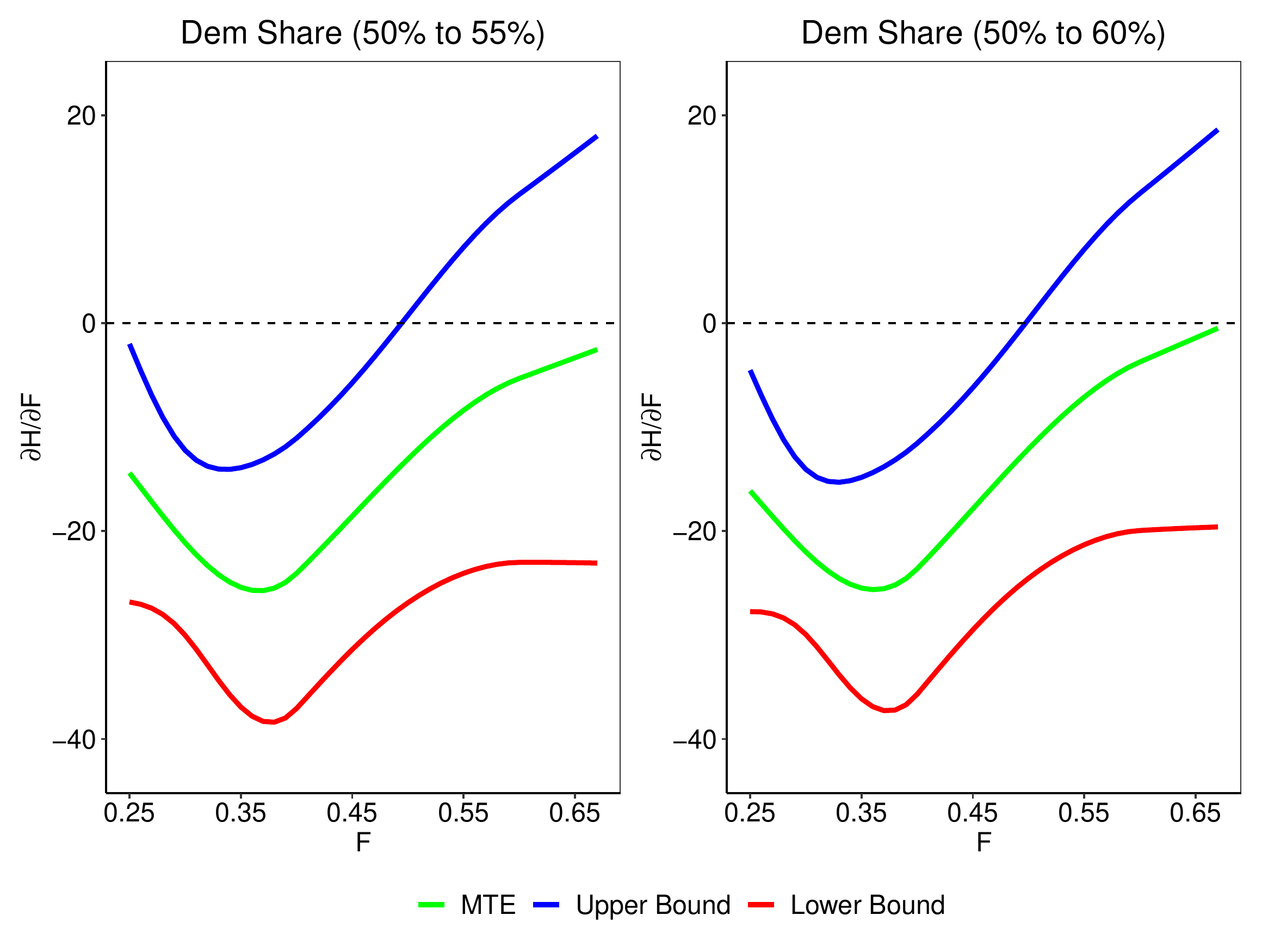"}
		\rule{\linewidth}{0in}
		\footnotesize
		\emph{Notes:} This figure plots the marginal treatment effect curves using a 5-knot cubic spline specification and a first stage probit model using political variables instead of the administrative barrier index. The first stage probit specifications correspond to those presented in Table \ref{tab:FSPPolVar}. Each specification omits the interaction between a Republican control state legislature and the Democratic gubernatorial candidates vote margin as the first stage instrument. Upper and lower bounds are generated from a block bootstrap with replacement at the state level using 500 samples. 
	\end{minipage}
\end{figure}

\begin{figure}[ht]	
	\caption{Marginal Labor Supply Curves for Different Natural Cubic Splines---Simple Average}\label{fig:MTEKnotCompSimpAvg}
	\hspace*{-2.0cm}
	\centering
	\begin{minipage}{\linewidth}
		\includegraphics[scale=0.80]{"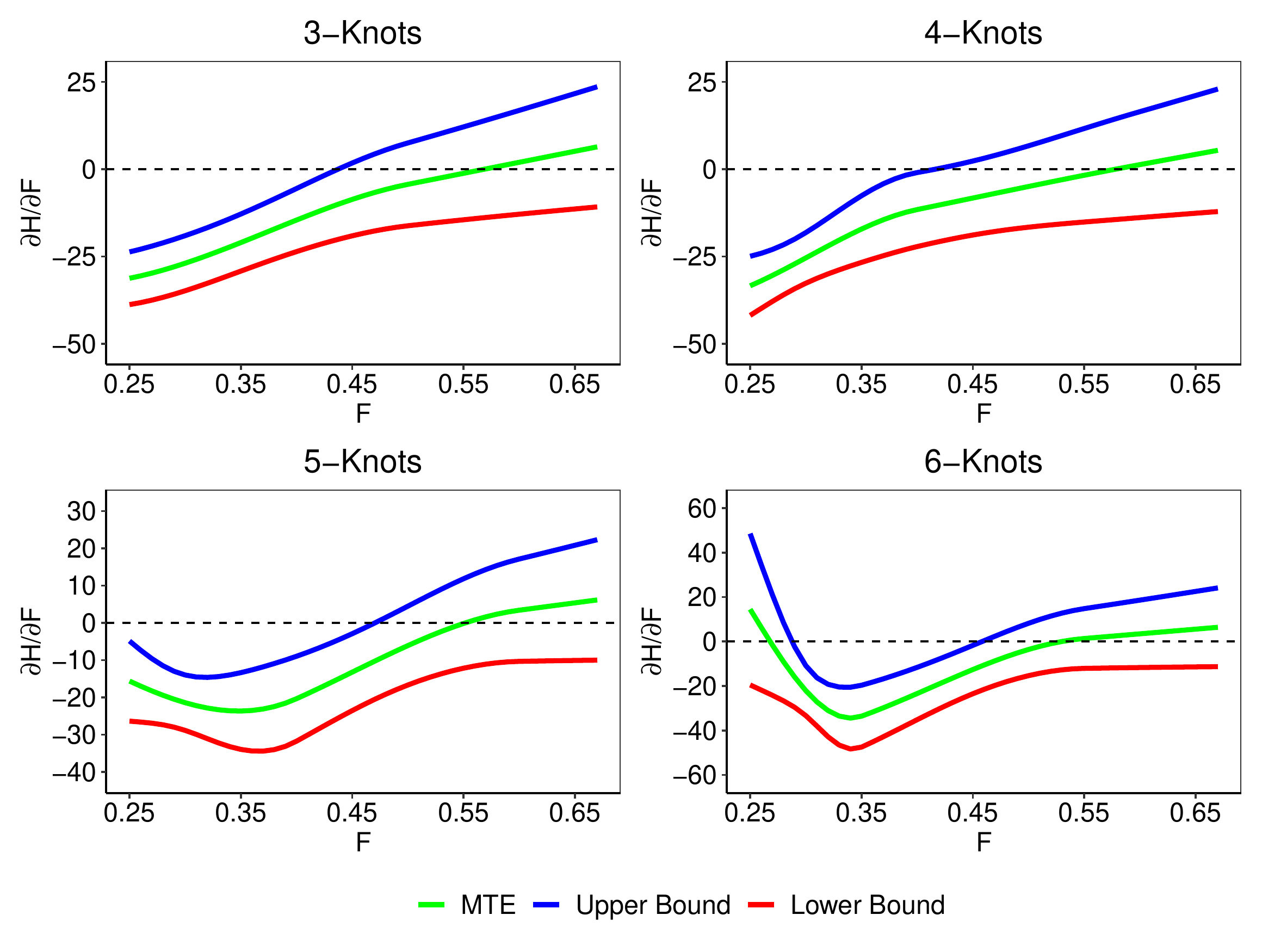"}
		\rule{\linewidth}{0in}
		\footnotesize
		\emph{Notes:} This figure plots the marginal treatment effect curves using different cubic spline specifications. All specifications use a first stage probit model with the simple average of the log of the AFDC administrative barriers and interactions with $N$, $G$, and $W(1-t)$. Upper and lower bounds are generated from a block bootstrap with replacement at the state level using 500 samples.  
	\end{minipage}
\end{figure}

\begin{figure}[ht]	
	\caption{Marginal Labor Supply Curves Using Selection Bias Adjusted Wage}\label{fig:MTEHeck}
	\hspace*{-2.0cm}
	\centering
	\begin{minipage}{\linewidth}
		\includegraphics[scale=0.80]{"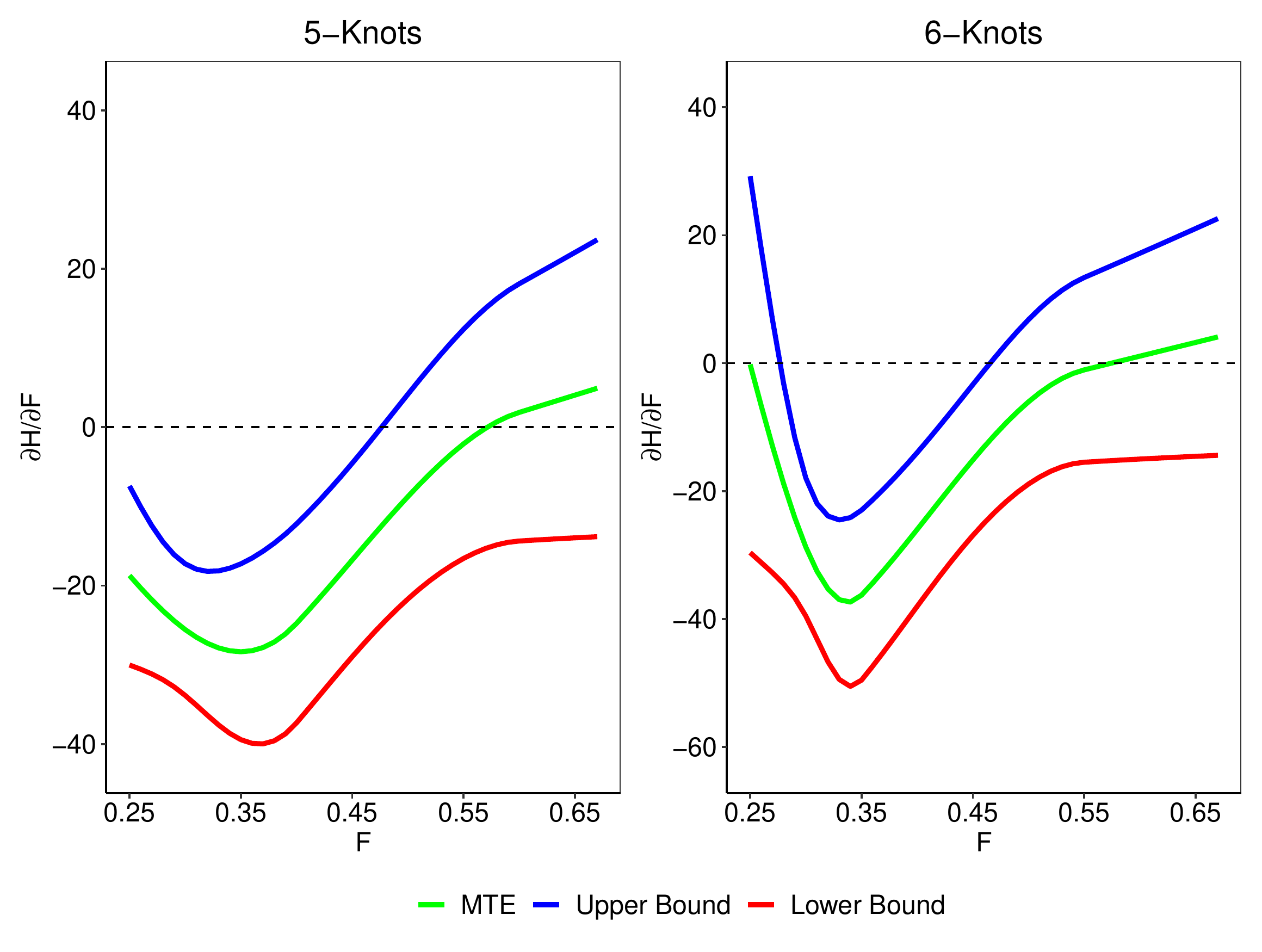"}
		\rule{\linewidth}{0in}
		\footnotesize
		\emph{Notes:} This figure plots the marginal treatment effect curves using 5 and 6-knot cubic spline specifications. All specifications use a first stage probit model with the simple average of the log of the AFDC administrative barriers and interactions with $N$, $G$, and $W(1-t)$. Wage variables values are predicted from the selection-bias adjusted model in Table \ref{tab:WageEqEst}. Upper and lower bounds are generated from a block bootstrap with replacement at the state level using 500 samples. 
	\end{minipage}
\end{figure}

Figure \ref{fig:MTEPolVars} shows the estimated MTE curves using these close election variables as instruments (i.e., including in the hours equations all the usual demographics plus the other variables in the Table \ref{tab:FSPPolVar} AFDC probits).  The MTE curves have approximately the same shape and locations of 95 percent confidence intervals as in our main results.  This provides further support for the main findings of the analysis.

\singlespacing\section{Marginal Labor Supply Disincentives for Three Major AFDC Reforms}
\doublespacing
\hspace{5ex} The AFDC program has experienced three major changes in the tax rate on benefits over its history.  From its creation in 1935 to 1967, the nominal tax rate was 100 percent.  This high tax rate was the subject of well-known criticisms of the program by \cite{Fried1962}, \cite{Lamp1965}, and \cite{Tobin1966} for its resulting work disincentives.  In 1967, Congress lowered the tax rate to 67 percent to provide work incentives to AFDC participants.  However, the Reagan Administration, in its early days in 1981, based on a prior reform in California when Reagan was Governor, concluded that low tax rates just increased the caseload and hence costs without any significant work incentives.  At the Administration's recommendation, Congress raised the tax rate in the program back to 100 percent.  A reversal of this decision took place in 1996, when major welfare program legislation transformed the AFDC program into a more pro-work program with work requirements and time limits.   As part of that reform, states were allowed to set their own tax rates rather than have them federally mandated, and most states chose to implement major reductions.  On average, the tax rate after the reform was approximately 50 percent.

A simple model of labor supply responses without much heterogeneity would predict that the 1981 tax rate increase would just reverse the labor supply effects of the 1967 tax reduction, and that the 1996 reduction would have effects similar to those of the 1967 reduction, although presumably slightly larger given the larger magnitude of the reduction.  However, the participation rate in the program was very different in the three reform years.  The rate was modest in 1967, around 0.36, but rose in the late 1960s and early 1970s before leveling off (\cite{Moff1992}).  By 1981, the participation rate was just over 0.50.  In the 1980s, the participation rate began to decline, reaching the 0.37 level in our 1988--1992 data but then rising again in the early 1990s. By 1996, the participation rate had risen back to 0.40 (\cite{Ziliak2016}).  Because marginal labor supply effects differ depending on the participation rate, marginal labor supply effects should have therefore been different at each of these historical periods.

In addition to differences in tax rates and participation rates, real guarantees were different in the three years.  Guarantees were very high in the 1960s and in 1967 in particular but, over the latter half of the 1970s and early 1980s, they were allowed to fall in real terms as state legislatures failed to raise the nominal amounts sufficiently to offset inflation.  By 1981, guarantees were 30 percent lower than they had been in 1967.  But over the early 1990s, states began raising guarantee levels again and, by 1996, they had reached a level about halfway between their 1967 high level and their 1981 low level.  Thus guarantee levels were also different in the different years, as were the initial tax rates at the time the tax-rate reforms took place.  Since the model shows that marginal effects depend on the initial levels of tax rates and guarantees, and since those affect the composition of the recipient population at the time of reform and therefore who is on the margin, marginal labor supply effects could also differ across periods for this reason.

We estimate the marginal labor supply disincentives at each of these three reform dates under the assumption that the model we have estimated in the late 1980s and 1990s was applicable to those periods.  This is obviously a strong assumption and the consequent conditional nature of these calculations must be understood.  To estimate the effects of these factors, we obtained Current Population Survey (CPS) files for 1967, 1981, 1988--1992, and 1996.  All demographic variables in the estimated participation and hours equations were constructed for each of those years from the CPS data (those for 1988--1992 are approximately the same as for the SIPP, but we choose to use the CPS to avoid any noncomparabilities across data sets).  The levels of $G$ and $t$ in the three reform years were also obtained.  Using our estimated participation equation from the SIPP data for 1988--1992 as reported above, the effects of changes in demographics as well as changes in the guarantees and tax rates on program participation between 1988--1992 and each of those other years on the participation rate could be calculated. Finally, using the fitted model of marginal labor supply effects reported above, those marginal effects could be computed for 1967, 1981, and 1996 at the participation rates existing in each of those years. 

The results are reported in Table \ref{tab:MTEreforms}.  In 1967, the demographics were not very different than those in 1988--1992 and only pushed the participation rate down by 2 percentage points, but the differing $G$ and $t$ values in 1967 pushed the participation rate upward by 7 percentage points relative to 1988--1992. These forces plus residual shifts moved the participation rate to 0.36.  At that participation rate and at the levels of $G$, $t$, and all demographics in 1967, the marginal individual had a labor supply effect of -27.9 hours, with a wide confidence interval but bounded away from zero.

But things were quite different in 1981, with the notable difference arising because the tax rate had risen to 1.0.  This change plus a small guarantee change led to a 16 percentage point decline in the participation rate.  Again, differences in demographics in 1981 and 1988--1992 had very little effect.  The participation rate, however, rose to 0.53, implying a large positive residual effect. At that participation rate and at the 1981 levels of $G$, $t$, and demographics, the marginal response was -9.2 hours per week and insignificantly different from zero.  Thus raising the tax rate back to its 1967 level did not have opposite marginal effects as had been its lowering because the participation rate and other factors made the marginal person different.

By 1996, guarantees had also risen by 20 percent from their 1988--1992 values.  The increase in the guarantee and a small change in $t$ pushed up the participation rate by 7 percentage points.  Changes in demographics again had little effect. At the 0.40 participation rate that obtained after including residual changes, the marginal response was about -26.4 hours, and hence had risen most of the way back to its 1967 level.\footnote{This simulation ignores all the structural changes in the program that occurred in 1996 and hence is only a hypothetical marginal response that would have occurred in the absence of those other reform elements.}

Simulations for marginal responses in years later than 1996 cannot be conducted with the model estimated in this paper because the program no longer took the simple form which the model represents.  However, participation rates in the program (now called TANF) are known to be approximately 10 to 15 percent.  Ignoring the other differences in the TANF and AFDC programs, this would imply that the hypothetical marginal labor supply response to an increase in participation at the current time would be insignificantly different from zero.

\section{Summary}

\hspace{5ex}This paper has provided a model and a reduced form estimation method for nonparametrically analyzing the marginal labor supply response in a classic transfer program of the textbook negative income tax type.  Applying the model to the Aid to Families with Dependent Children in the late 1980s and early 1990s and identifying marginal responses by variation in administrative barriers, the paper shows that marginal labor supply responses are non-montonic and quadratic, with the magnitude of the marginal response increasing as participation rates increase but eventually declining after participation rates pass an inflection point.  Marginal responses are insignificantly different from zero at low and high participation rates but negative and large in a middle range of such rates.  We show that traditional IV, which estimates a weighted average of marginal responses, gives a misleading picture of true marginal responses at different points of program expansion and contraction.  Using the estimates in a counterfactual exercise to quantify marginal responses at three historical years when major reforms of the program took place shows that marginal responses were different in each year, both because the demographic composition of the caseload was different, the level of the program parameters was different, and because a different fraction of the population was participating in the program. The largest marginal response was in 1967 when participation rates were fairly low and guarantees were fairly high.  The lowest marginal response was in 1981, when participation rates were high and guarantees were low and, in that year, a 95 percent confidence interval includes zero.  The marginal response in 1996 had risen back up almost to the 1967 level.

A number of obvious extensions of the analysis would be worthwhile.  One is to estimate a structural model which pins down the underlying parameters of a formally defined utility function whose parameters vary in the population.  That would allow a better analysis of counterfactuals than the method used here.   Another is to extend the static model to dynamic models where dynamics are introduced through intertemporal elasticities of labor supply, human capital, and preference persistence (\cite{ChanMoff2018}).   Yet another avenue for more model development is to add an analysis of inframarginal responses to transfer program reforms to the analysis of marginal responses, since any reform involving alteration of program parameters affects both.

There are also many programs of interest other than the simple negative-income-tax cash program type analyzed here.  The 1996 reform of the AFDC program introduced work requirements, time limits, and other features, which have been show to have had effects on average labor supply (\cite{Chan2013}).   Their marginal effects are likely to be quite different than those analyzed here because those reforms almost surely affected different portions of the labor supply preference distribution.  In addition, the participation rate in the program has dropped by 80 percent since those reforms, which surely affects who is on the margin of participation in the program.  The analysis of the marginal responses to in-kind transfers, which requires modeling the consumption of the subsidized good jointly with labor supply, is another obvious extension given the expansion of those types of transfers in the U.S. over the last 30 years.  

\newpage
\singlespacing
\bibliographystyle{chicago}
\bibliography{citations_master_2022-1-19}

\clearpage
\onehalfspacing

\section*{Tables} \label{sec:Tabs}
\addcontentsline{toc}{section}{Tables}

\begin{table}[!h]
\caption{\label{tab:AdminBars}Seven Administrative Barrier Variables}
\begin{center}
\begin{tabular}[t]{lrrrr}
\toprule
 & Mean & Std. Dev. & Min & Max\\
\midrule
Pct. ineligible in error & 1.7 & 0.8 & 0.3 & 4.7\\
Pct. hearings and appeals improperly denied & 2.0 & 1.5 & 0.4 & 5.8\\
Pct. cases elig. denied for non-grant reasons & 0.2 & 0.1 & 0.0 & 0.4\\
Pct. applications denied & 24.3 & 11.2 & 5.3 & 47.8\\
Pct. applications denied for procedural reasons & 14.1 & 8.9 & 1.3 & 34.6\\
Error rate in payment determination & 4.7 & 1.2 & 2.2 & 7.3\\
Error rate resulting in underpayment & 3.5 & 2.6 & 1.6 & 10.2\\
\bottomrule
\end{tabular}
\end{center}
\footnotesize \emph{Notes:} This table summarizes different administrative barriers for enrollment into the AFDC program from 1980--1992. The variables are averages over all years for each state.\\ \emph{Source:} Quarterly Public Assistance Statistics and unpublished data from the U.S. Department of Health and  Human Services.
\end{table}

\begin{table}[H]
\caption{\label{tab:ZonAFDCPart}Estimated Impact of Instruments on AFDC Participation}
\begin{center}
\resizebox{\linewidth}{!}{
\begin{tabular}[t]{lrrrcrr}
\toprule
\multicolumn{1}{c}{ } & \multicolumn{3}{c}{Probit Single Z} & \multicolumn{1}{c}{OLS Other Zs} & \multicolumn{2}{c}{Probit All Zs} \\
\cmidrule(l{3pt}r{3pt}){2-4} \cmidrule(l{3pt}r{3pt}){5-5} \cmidrule(l{3pt}r{3pt}){6-7}
Log Z & Est & SE & dy/dx & $R^2$ & Est & SE\\
\midrule
Pct. ineligible in error & -0.09 & 0.13 & -0.03 & 0.40 & -0.04 & 0.25\\
Pct. hearings and appeals improperly denied & 0.00 & 0.12 & 0.00 & 0.32 & -0.02 & 0.23\\
Pct. cases elig. denied for non-grant reasons & -0.07 & 0.13 & -0.02 & 0.39 & -0.05 & 0.20\\
Pct. applications denied & -0.10 & 0.18 & -0.03 & 0.81 & -0.17 & 0.54\\
Pct. applications denied for procedural reasons & -0.04 & 0.11 & -0.01 & 0.71 & 0.05 & 0.29\\
Error rate in payment determination & -0.07 & 0.24 & -0.02 & 0.10 & 0.01 & 0.40\\
Error rate resulting in underpayment & 0.00 & 0.18 & 0.00 & 0.61 & -0.08 & 0.41\\
\midrule
\midrule
Inverse Variance Weighted Average & -0.25 & 0.37 & -0.08 & 1.00 &  & \\
Simple Average & -0.22 & 0.31 & -0.07 & 1.00 &  & \\
\bottomrule
\end{tabular}
}
\end{center}
\footnotesize \emph{Notes:} This table reports the impacts of the log of the administrative barrier instruments on AFDC participation. The first three columns report the coefficient estimate, standard error, and marginal effect from a probit model using the logged instrument in the rows. The fourth column reports the $R^2$ from an OLS regression of the logged administrative barrier instrument in each row onto the other six. The fifth and sixth columns report the coefficient estimate and standard error for the logged variable in each row when the other six are included in the equation. The last two rows in the table show the impact of the two aggregated barrier indices discussed in the text.  All equations include $\log(W)$, $\log[W(1-t)]$, $\log(G)$, $\log(N+10)$ (because some observations have $N=0$), age, black, family size, the number of children less than 6, the state unemployment rate, three regional dummies (a fourth is omitted), and the Food Stamp guarantee. The estimated coefficients for the full specification for the two barrier index equations are shown in Appendix Table \ref{tab:FSPEstDetail}. Standard errors in columns two and six are generated from a block bootstrap with replacement at the state level using 500 samples.
\end{table}


\begin{table}[H]
\caption{First Stage Probit Budget Constraint Interaction Comparison}
\begin{center}
\begin{scriptsize}
\begin{tabular}{l c c c c c c c c c c}
\toprule
 & \multicolumn{5}{c}{Inverse Variance Weighted Average} & \multicolumn{5}{c}{Simple Average} \\
\cmidrule(lr){2-6} \cmidrule(lr){7-11}
 & (1) & (2) & (3) & (4) & (5) & (6) & (7) & (8) & (9) & (10) \\
\midrule
Log Z          & $-0.25$   & $-0.15$       & $0.25$        & $1.92^{**}$   & $1.90^{*}$    & $-0.22$   & $-0.11$       & $0.28$        & $1.72^{***}$  & $1.90^{*}$    \\
               & $(0.37)$  & $(0.38)$      & $(0.37)$      & $(0.78)$      & $(0.97)$      & $(0.31)$  & $(0.33)$      & $(0.31)$      & $(0.61)$      & $(0.97)$      \\
Log Z*N        &           & $-0.01^{***}$ & $-0.01^{***}$ & $-0.01^{***}$ & $-0.01^{***}$ &           & $-0.01^{***}$ & $-0.01^{***}$ & $-0.01^{***}$ & $-0.01^{***}$ \\
               &           & $(0.00)$      & $(0.00)$      & $(0.00)$      & $(0.00)$      &           & $(0.00)$      & $(0.00)$      & $(0.00)$      & $(0.00)$      \\
Log Z*G        &           &               & $-0.05^{*}$   & $-0.05^{*}$   & $-0.05^{*}$   &           &               & $-0.05^{*}$   & $-0.05$       & $-0.05^{*}$   \\
               &           &               & $(0.03)$      & $(0.03)$      & $(0.03)$      &           &               & $(0.03)$      & $(0.03)$      & $(0.03)$      \\
Log Z*$W(1-t)$ &           &               &               & $-0.49^{**}$  & $-0.50^{*}$   &           &               &               & $-0.42^{***}$ & $-0.50^{*}$   \\
               &           &               &               & $(0.21)$      & $(0.30)$      &           &               &               & $(0.16)$      & $(0.30)$      \\
Log Z*W        &           &               &               &               & $0.01$        &           &               &               &               & $0.01$        \\
               &           &               &               &               & $(0.24)$      &           &               &               &               & $(0.24)$      \\
\midrule
Obs            & $3381$ & $3381$     & $3381$     & $3381$     & $3381$     & $3381$ & $3381$     & $3381$     & $3381$     & $3381$     \\
\bottomrule
\end{tabular}
\end{scriptsize}
\label{tab:FSPBCInts}
\end{center}
\end{table}

\begin{table}[H]
	\vspace*{-0.7cm}
	\noindent\scriptsize * $(p<0.1)$, ** $(p<0.05)$, *** $(p<0.01)$. \\	\emph{Notes:} Standard errors in parentheses are generated from a block bootstrap with replacement at the state level using 500 samples. The presented coefficient estimates are from the AFDC participation probit using aggregates of the log of the administrative barriers and interactions with individual budget constraint variables. The aggregate used is denoted above the specification number. All specifications include $\log(W)$, $\log[W(1-t)]$, $\log(G)$, $\log(N+10)$ (because some observations have $N=0$), age, black, family size, the number of children less than 6, the state unemployment rate, three regional dummies (a fourth is omitted), and the Food Stamp guarantee.
\end{table}

\begin{table}[!h]
\caption{\label{tab:OPHIBalSum}Share of Covariates Balanced Given the Generalized Propensity Score}
\begin{center}
\begin{tabular}[t]{ccccc}
\toprule
\multicolumn{1}{c}{ } & \multicolumn{2}{c}{Inv. Var. Wtg. Avg.} & \multicolumn{2}{c}{Simple Avg.} \\
\cmidrule(l{3pt}r{3pt}){2-3} \cmidrule(l{3pt}r{3pt}){4-5}
Specification & Unadj. & GPS Adj. & Unadj. & GPS Adj.\\
\midrule
10 & 0.308 & 0.692 & 0.385 & 0.769\\
11 & 0.462 & 0.769 & 0.154 & 0.615\\
15 & 0.385 & 0.769 & 0.462 & 0.615\\
16 & 0.308 & 0.538 & 0.308 & 0.923\\
\bottomrule
\end{tabular}
\end{center}
\footnotesize \emph{Notes:} This table reports the share of covariates that are balanced following the generalized propensity score (GPS) adjustment from \cite{HirImb2004}. Balance is measured from the t-test statistics for the equality of means for observations above and below the median value of the aggregates of the log of the AFDC administrative barriers. The unadjusted columns report the share of balanced test statistics that do not adjust for the GPS. The GPS adjusted columns report the share of balanced test statistics following the adjustment procedure. The GPS is generated from an ordered probit model. All specifications include the AFDC participation probit covariates and different polynomials and interactions. Specification (10) includes cubes of the continuous covariates and interactions with $\log[W(1-t)]$. Specification (11) includes squares of the continuous covariates and interactions with black and $\log[W(1-t)]$. Specification (15) is the same as (10) but adds interactions with black and $\log(G)$. Specification (16) includes the continuous covariates squared and interactions with black, the food stamp guarantee, and the state unemployment rate.
\end{table}

\begin{table}[!h]
\caption{\label{tab:pFSum}F-Statistics by Different Participation Probability Ranges}
\begin{center}
\resizebox{\linewidth}{!}{
\begin{tabular}[t]{ccccccccc}
\toprule
\multicolumn{1}{c}{ } & \multicolumn{4}{c}{Inv. Var. Wtg. Avg.} & \multicolumn{4}{c}{Simple Avg.} \\
\cmidrule(l{3pt}r{3pt}){2-5} \cmidrule(l{3pt}r{3pt}){6-9}
Part. Prob. Range & None & N & N \& G & N, G, \& W(1-t) & None & N & N \& G & N, G, \& W(1-t)\\
\midrule
0.00--0.33 & 1.43 & 1.84 & -0.04 & 0.71 & 1.49 & 2.41 & 0.37 & 1.29\\
0.33--0.66 & 1.32 & 12.49 & 10.71 & 8.63 & 2.02 & 11.32 & 10.32 & 7.78\\
0.66--1.00 & 0.71 & 1.65 & 2.78 & 3.12 & -0.23 & 1.23 & 2.13 & 2.30\\
\midrule
0.00--0.25 & 1.63 & 0.61 & -0.04 & 0.22 & 1.69 & 0.80 & 0.03 & 0.87\\
0.25--0.50 & 1.13 & 10.74 & 7.48 & 6.72 & 1.21 & 10.36 & 7.69 & 6.07\\
0.50--0.75 & 0.02 & 3.87 & 4.68 & 3.95 & 0.09 & 3.14 & 4.14 & 3.27\\
0.75--1.00 & 0.67 & 0.76 & 1.32 & 1.56 & 0.29 & 0.65 & 0.97 & 1.16\\
\bottomrule
\end{tabular}
}
\end{center}
\footnotesize \emph{Notes:} This table reports the F-statistics within different participation probability ranges. To calculate the F-statistic within a specific range of $\hat{F}$, define $RSS(q)$ as the residual sum of squares, equal to the sum of $[P-\hat{F}]^2$ taken over all observations in the range. The F-stat is calculated as (1) the difference in $RSS(q)$ for the restricted model excluding the instruments and the unrestricted model $RSS(q)$ including the instruments divided by the d.o.f., divided by (2) the residual variance computed over all observations in the sample, using $\hat{F}$ from the restricted model. Participation probabilities come from the probit models in Table 
\ref{tab:FSPBCInts} for the inverse variance weighted average aggregate of the log of the administrative barriers. Column headings note which variables are interacted with the aggregate.
\end{table}

\begin{table}[!h]
\caption{\label{tab:GCVSum}Generalized Cross-Validation Statistics by First Stage Probit and $g$ Cubic Spline Specifications}
\begin{center}
\begin{tabular}[t]{llcccc}
\toprule
\multicolumn{2}{c}{ } & \multicolumn{4}{c}{Knots} \\
\cmidrule(l{3pt}r{3pt}){3-6}
 & FSP Interactions & 3 & 4 & 5 & 6\\
\midrule
 & $N$ & 325.30 & 325.34 & 324.84 & 324.66\\
 & $N$ and $G$ & 325.16 & 325.23 & 324.55 & 324.52\\
\multirow{-3}{*}{\raggedright\arraybackslash Inv. Var. Wtg. Avg.} & $N$, $G$, and $W(1-t)$ & 324.97 & 325.12 & 324.65 & 324.76\\
\midrule
 & $N$ & 325.24 & 325.26 & 324.48 & 324.29\\
 & $N$ and $G$ & 325.02 & 325.08 & 324.12 & 324.00\\
\multirow{-3}{*}{\raggedright\arraybackslash Simple Avg.} & $N$, $G$, and $W(1-t)$ & 324.78 & 324.88 & 324.15 & 324.16\\
\bottomrule
\end{tabular}
\end{center}
\footnotesize \emph{Notes:} This table reports the generalized cross-validation (GCV) statistic for different first stage participation probit and $g$ cubic spline specifications in the hours equation. All first stage probit specifications include $\log(W)$, $\log[W(1-t)]$, $\log(G)$, $\log(N+10)$ (because some observations have $N=0$), age, black, family size, the number of children less than 6, the state unemployment rate, three regional dummies (a fourth is omitted), and the Food Stamp guarantee. The aggregate of the logged AFDC administrative barrier aggregate that is included is denoted in the first column. Column headings denote the number of knots used in the hours equation cubic spline.
\end{table}

\clearpage


\begin{center}
\begin{small}
\begin{longtable}[h]{l c c c}
\caption{Estimates of Hours Equation $\lambda$ and $g$ Coefficients with Five-Knot $g$ Spline}
\label{tab:HoursEqSplineInt}\\
\toprule
 & (1) & (2) & (3) \\
\midrule
\endfirsthead
\toprule
 & (1) & (2) & (3) \\
\midrule
\endhead
\bottomrule
\endfoot
\bottomrule
\endlastfoot
$\pmb{g}$                        &                &               &               \\
                                 &                &               &               \\
\quad Constant*10                & $20.99^{**}$   & $19.03^{**}$  & $19.69^{**}$  \\
                                 & $(8.35)$       & $(8.49)$      & $(8.59)$      \\
\quad $\hat{F}$*100              & $-17.27^{***}$ & $-16.15^{**}$ & $-16.57^{**}$ \\
                                 & $(6.55)$       & $(6.70)$      & $(6.63)$      \\
\quad S3*1000                    & $41.31^{**}$   & $39.30^{**}$  & $39.87^{**}$  \\
                                 & $(17.69)$      & $(17.89)$     & $(17.80)$     \\
\quad S4*1000                    & $-56.68^{**}$  & $-54.01^{**}$ & $-54.71^{**}$ \\
                                 & $(24.79)$      & $(25.00)$     & $(24.93)$     \\
\quad S5*1000                    & $15.62^{**}$   & $14.96^{**}$  & $15.07^{**}$  \\
                                 & $(7.34)$       & $(7.34)$      & $(7.39)$      \\
$\pmb{\lambda^1}$                &                &               &               \\
                                 &                &               &               \\
\quad Log $\hat{W}$           & $-8.32$        & $-1.75$       & $-16.50$      \\
                                 & $(21.40)$      & $(52.72)$     & $(23.43)$     \\
\quad N                          & $0.17$         & $0.32$        & $0.18$        \\
                                 & $(0.13)$       & $(0.27)$      & $(0.15)$      \\
\quad Log G                      & $-3.85$        & $-4.05$       & $-2.49$       \\
                                 & $(7.65)$       & $(17.38)$     & $(8.27)$      \\
\quad Log $\hat{W}(1-t)$         & $-4.95$        & $5.81$        & $-5.68$       \\
                                 & $(12.92)$      & $(32.70)$     & $(14.12)$     \\
\quad Age                      &                &               & $0.49$        \\
                                 &                &               & $(0.36)$      \\
\quad Black                    &                &               & $3.88$        \\
                                 &                &               & $(3.69)$      \\
\textbf{Interactions}            &                &               &               \\
                                 &                &               &               \\
\quad Log $\hat{W}*\hat{F}$      &                & $-3.59$       &               \\
                                 &                & $(60.51)$     &               \\
\quad N*$\hat{F}$                 &                & $-0.25$       &               \\
                                 &                & $(0.37)$      &               \\
\quad Log G*$\hat{F}$            &                & $-0.85$       &               \\
                                 &                & $(19.29)$     &               \\
\quad Log $\hat{W}(1-t)*\hat{F}$ &                & $-19.92$      &               \\
                                 &                & $(44.57)$     &               \\
$\pmb{\beta}$                    &                &               &               \\
                                 &                &               &               \\
\quad Log $\hat{W}$            & $24.69^{**}$   & $23.67^{**}$  & $28.12^{*}$   \\
                                 & $(10.73)$      & $(11.56)$     & $(14.38)$     \\
\quad Log (N+10)                 & $-1.53$        & $-1.87^{*}$   & $-1.65^{*}$   \\
                                 & $(0.96)$       & $(1.05)$      & $(1.00)$      \\
\quad Age                      & $-0.10$        & $-0.11$       & $-0.27$       \\
                                 & $(0.11)$       & $(0.12)$      & $(0.23)$      \\
\quad Black                    & $-0.84$        & $-0.88$       & $-2.31$       \\
                                 & $(0.98)$       & $(1.08)$      & $(1.99)$      \\
\quad Family Size                & $-0.83$        & $-0.84$       & $-1.03^{*}$   \\
                                 & $(0.61)$       & $(0.67)$      & $(0.59)$      \\
\quad Number of Children $<$ 6   & $-2.10^{**}$   & $-2.23^{**}$  & $-1.94^{*}$   \\
                                 & $(0.96)$       & $(1.12)$      & $(1.03)$      \\
\quad Food Stamp Guarantee       & $-11.26$       & $-11.55$      & $-10.51$      \\
                                 & $(15.11)$      & $(16.44)$     & $(15.72)$     \\
\quad Unemployment Rate          & $-0.69^{**}$   & $-0.73^{**}$  & $-0.68^{**}$  \\
                                 & $(0.30)$       & $(0.33)$      & $(0.32)$      \\
\quad Northeast                  & $-9.44^{**}$   & $-9.60^{**}$  & $-10.10^{**}$ \\
                                 & $(4.33)$       & $(4.69)$      & $(4.85)$      \\
\quad Midwest                    & $-2.14$        & $-2.27$       & $-2.62$       \\
                                 & $(3.26)$       & $(3.46)$      & $(3.54)$      \\
\quad West                       & $-5.29$        & $-5.43$       & $-5.86$       \\
                                 & $(3.94)$       & $(4.19)$      & $(4.41)$      \\
\quad Constant                   & $9.89$         & $14.38$       & $11.06$       \\
                                 & $(24.84)$      & $(25.80)$     & $(27.28)$     \\
\midrule
GCV                              & $324.9$       & $325.6$      & $324.6$      \\
Obs                              & $3381$      & $3381$     & $3381$     \\
\end{longtable}
\end{small}
\end{center}

\begin{table}[H]
	\vspace*{-5.0cm}
	\noindent\scriptsize * $(p<0.1)$, ** $(p<0.05)$, *** $(p<0.01)$. \\	\emph{Notes:} Standard errors in parentheses are generated from a block bootstrap with replacement at the state level using 500 samples. $\hat{F}$ is generated from a first stage probit specification that includes the inverse variance weighted log of the AFDC administrative barriers and interactions with $N$, $G$, and $W(1-t)$. \\ 1: Variables expressed as deviations from means.
\end{table}
\clearpage

\begin{table}[!h]
\caption{\label{tab:ShareHbyP}Percent Working 0, 20, and 40 Hours by Welfare Participation Status}
\begin{center}
\begin{tabular}[t]{cccc}
\toprule
 & H = 0 & H = 20 & H = 40\\
\midrule
All & 42.7 & 12.0 & 45.3\\
P = 0 & 19.0 & 13.3 & 67.7\\
P = 1 & 83.7 & 9.8 & 6.5\\
\bottomrule
\end{tabular}
\end{center}
\footnotesize \emph{Notes:} This table reports the share of women in our analysis sample that are not working, working part-time, and working full-time. The first row reports this tabulation for the entire sample, the second row reports this for women not on welfare, and the third row reports this women on welfare.
\end{table}

\begin{table}[!h]
\caption{\label{tab:VarMeansbyFZ}Variable Means for Observations in Quintiles of $\hat{F}$ Distribution}
\begin{center}
\begin{tabular}[t]{lccccc}
\toprule
 & 1st Quintile & 2nd Quintile & 3rd Quintile & 4th Quintile & 5th Quintile\\
\midrule
Hourly wage & 6.38 & 6.03 & 5.81 & 5.78 & 5.59\\
Weekly non-labor inc & 8.80 & 8.57 & 5.97 & 2.21 & 2.35\\
Black & 0.30 & 0.38 & 0.40 & 0.42 & 0.52\\
Age & 35.01 & 32.32 & 30.22 & 28.88 & 28.75\\
Children $<$ 6 & 0.40 & 0.61 & 0.71 & 0.91 & 1.33\\
\bottomrule
\end{tabular}
\end{center}
\footnotesize \emph{Notes:} This table reports variable means within quintiles of the center of the $\hat{F}$ distribution (0.25 to 0.66). The cutoffs for these quintiles are approximately 0.32, 0.39, 0.46, and 0.53. $\hat{F}$ is generated from a probit model using the inverse variance weighted index of the log of the AFDC administrative barrier variables and interactions with $N$, $G$, and $W(1-t)$.
\end{table}

\clearpage


\begin{table}[h]
\caption{First Stage Estimates Using 1992 Law Change}\label{tab:LogZTSMicroReg}
\begin{center}
\begin{tabular}{l c c c}
\toprule
 & \multicolumn{1}{c}{OLS} & \multicolumn{2}{c}{Probit} \\
\cmidrule(lr){2-2} \cmidrule(lr){3-4}
 & Z Index & AFDC & AFDC \\
\midrule
Time Trend                          & $0.08^{***}$ & $0.14$      &           \\
                                    & $(0.02)$     & $(0.17)$    &           \\
[1em]
1992 Flag                           & $0.06^{*}$   & $-0.22^{*}$ &           \\
                                    & $(0.03)$     & $(0.12)$    &           \\
[1em]
Log $\hat{Z}$ (Inv. Var. Wgt. Avg.) &              &             & $-0.25$   \\
                                    &              &             & $(0.32)$  \\
[1em]
1992 Residual (Inv. Var. Wgt. Avg.) &              &             & $-0.46$   \\
                                    &              &             & $(0.30)$  \\
[1em]
Probit Controls     &                     &  \checkmark         &  \checkmark         \\
\midrule
Obs                                 & $3381$    & $3381$   & $3381$ \\
\bottomrule
\end{tabular}
\end{center}
\end{table}

\begin{table}[H]
	\vspace*{-0.7cm}
	\noindent\footnotesize * $(p<0.1)$, ** $(p<0.05)$, *** $(p<0.01)$. \\ \emph{Notes:} Standard errors in parentheses are generated from a block bootstrap with replacement at the state level using 500 samples. The first column reports the OLS estimates of the log inverse variance weighted index onto a time trend and indicator for 1992. The second and third columns report AFDC probit estimate using listed variables in place of the administrative barriers index. ``Log $\hat{Z}$" is the predicted value from state-year level OLS regressions of the log administrative barrier index onto a time trend using a sample from 1980--1993. ``1992 Residual" is the residual between Log $\hat{Z}$ and Log $Z$ in 1992. 
\end{table}
\clearpage


\begin{table}[h]
\caption{First Stage Estimates Using Political RD}
\begin{center}
\begin{small}
\begin{tabular}{l c c c c}
\toprule
 & \multicolumn{2}{c}{OLS} & \multicolumn{2}{c}{Probit} \\
\cmidrule(lr){2-3} \cmidrule(lr){4-5}
 & Z Index & Z Index & AFDC & AFDC \\
\midrule
\textbf{Elections}                              &              &              &               &               \\
[1em]
\quad Dem Gov Share                             & $-0.31$      & $-0.28$      & $0.44$        & $0.53$        \\
                                                & $(0.26)$     & $(0.24)$     & $(0.42)$      & $(0.42)$      \\
[1em]
\quad Dem Gov Share (50\% to 55\%)            & $-0.07$      &              & $-0.08$       &               \\
                                                & $(0.07)$     &              & $(0.11)$      &               \\
[1em]
\quad Dem Gov Share (50\% to 60\%)            &              & $-0.09$      &               & $-0.03$       \\
                                                &              & $(0.06)$     &               & $(0.12)$      \\
[1em]                                                
\textbf{State Legislature}                      &              &              &               &               \\
[1em]
\quad Republican                                & $-0.03$      & $0.01$       & $-0.03$       & $0.09$        \\
                                                & $(0.11)$     & $(0.10)$     & $(0.19)$      & $(0.20)$      \\
[1em]
\quad Split                                     & $0.25^{***}$ & $0.24^{***}$ & $-0.12$       & $-0.13$       \\
                                                & $(0.09)$     & $(0.09)$     & $(0.15)$      & $(0.16)$      \\
[1em]
\textbf{Interactions}                           &              &              &               &               \\
[1em]
\quad Republican$\times$Dem Share(50\% to 55\%) & $0.27^{**}$  &              & $-0.65^{***}$ &               \\
                                                & $(0.11)$     &              & $(0.20)$      &               \\
[1em]
\quad Republican$\times$Dem Share(50\% to 60\%) &              & $0.11$       &               & $-0.89^{***}$ \\
                                                &              & $(0.16)$     &               & $(0.32)$      \\
[1em]
Probit Controls     &  \checkmark         &  \checkmark         &  \checkmark         &  \checkmark         \\
\midrule
Obs                                             & $3169$    & $3169$    & $3169$     & $3169$     \\
$R^2$                                           & $0.52$       & $0.52$       & $$            & $$            \\
\bottomrule
\end{tabular}
\end{small}
\label{tab:FSPPolVar}
\end{center}
\end{table}

\begin{table}[H]
	\vspace*{-0.7cm}
	\noindent\footnotesize * $(p<0.1)$, ** $(p<0.05)$, *** $(p<0.01)$. \\ \emph{Notes:} Standard errors in parentheses are generated from a block bootstrap with replacement at the state level using 500 samples. The first column reports the OLS estimates of the log inverse variance weighted index onto a time trend and indicator for 1992. The second and third columns report AFDC probit estimate using listed variables in place of the administrative barriers index. ``Log $\hat{Z}$" is the predicted value from state-year level OLS regressions of the log administrative barrier index onto a time trend using a sample from 1980--1993. ``1992 Residual" is the residual between Log $\hat{Z}$ and Log $Z$ in 1992. 
\end{table}
\clearpage

\begin{table}[htbp]
  \centering
  \caption{Marginal Labor Supply Effects at Three Historical Reforms}\label{tab:MTEreforms}%
    \begin{tabular}{rrr}
    \toprule
      &   &  \\
    \midrule
        \textbf{1967} &   &  \\
        & \multicolumn{1}{l}{Participation rate} & \multicolumn{1}{c}{0.36} \\
        & \multicolumn{1}{l}{Difference due to demographics} & \multicolumn{1}{c}{-0.02} \\
        & \multicolumn{1}{l}{Difference due to $G$ and $t$} & \multicolumn{1}{c}{+0.07} \\
        &   &  \\
        & \multicolumn{1}{l}{Marginal labor supply effect} & \multicolumn{1}{c}{-27.9} \\
        &   & \multicolumn{1}{c}{(-39.15,-16.65)} \\
        &   &  \\
        \textbf{1981} &   &  \\
        & \multicolumn{1}{l}{Participation rate} & \multicolumn{1}{c}{0.53} \\
        & \multicolumn{1}{l}{Difference due to demographics} & \multicolumn{1}{c}{-0.02} \\
        & \multicolumn{1}{l}{Difference due to $G$ and $t$} & \multicolumn{1}{c}{-0.16} \\
        &   &  \\
        & \multicolumn{1}{l}{Marginal labor supply effect} & \multicolumn{1}{c}{-9.2} \\
        &   & \multicolumn{1}{c}{(-20.56,4.38)} \\
        &   &  \\
        \textbf{1996} &   &  \\
        & \multicolumn{1}{l}{Participation rate} & \multicolumn{1}{c}{0.40} \\
        & \multicolumn{1}{l}{Difference due to demographics} & \multicolumn{1}{c}{+0.00} \\
        & \multicolumn{1}{l}{Difference due to $G$ and $t$} & \multicolumn{1}{c}{+0.07} \\
        &   &  \\
        & \multicolumn{1}{l}{Marginal labor supply effect} & \multicolumn{1}{c}{-26.4} \\
        &   & \multicolumn{1}{c}{(-37.08,-13.43)} \\
        &   &  \\
    \bottomrule
    \end{tabular}%
\end{table}%

\begin{table}[H]
	\vspace*{-0.5cm}
	\noindent\footnotesize \emph{Notes:} This table calculates the marginal labor supply responses in three periods when the AFDC program was reformed. The calculation uses parameter estimates from our preferred model specification---first stage participation probit using the inverse variance weighted log of the AFDC administrative barriers and interactions with $N$, $G$, and $W(1-t)$ and a 5-knot cubic spline in the hours equation. Data for the historic periods in the table come from the CPS.
\end{table}
\clearpage

\section*{Figures} \label{sec:Figs}
\addcontentsline{toc}{section}{Figures}

\clearpage
\appendix

\section{Additional Tables}\label{app:addtabs}

\begin{table}[!h]
\caption{\label{tab:VariableMeans}Means of Variables Used in the Analysis}
\begin{center}
\begin{tabular}[t]{lrrr}
\toprule
 & Full Sample & P=1 & P=0\\
\midrule
Weekly H & 21.38 & 4.48 & 31.14\\
P & 0.37 & 1.00 & 0.00\\
Log $\hat{W}$ & 1.78 & 1.74 & 1.81\\
Log (N+10) & 2.97 & 2.58 & 3.19\\
Log G & -2.49 & -2.38 & -2.55\\
Log $\hat{W}$(1-t) & 1.27 & 1.22 & 1.30\\
Age & 32.48 & 30.27 & 33.75\\
Black & 0.34 & 0.41 & 0.30\\
Education & 10.89 & 10.49 & 11.13\\
Family size & 3.09 & 3.37 & 2.94\\
No. Children $<$ 6 & 0.79 & 1.14 & 0.58\\
Food Stamp Guarantee & 0.78 & 0.78 & 0.78\\
Unemployment rate & 6.35 & 6.44 & 6.30\\
Northeast & 0.28 & 0.28 & 0.28\\
Midwest & 0.27 & 0.27 & 0.26\\
West & 0.22 & 0.25 & 0.20\\
State Percent Services & 27.67 & 27.94 & 27.52\\
State Percent Manufacturing & 15.39 & 15.31 & 15.43\\
State Percent Urban & 76.26 & 77.49 & 75.55\\
Obs & 3,381 & 1,238 & 2,143\\
\bottomrule
\end{tabular}
\end{center}
\footnotesize \emph{Notes:} This table reports the means of variables used in our analysis. The sample is composed of single mothers aged 25--55 with a high school education or less with total assets less than \$1,500 a week and non-transferable non-labor income less than \$1,000 a week drawn from 1988-1992 SIPP interviews. All dollar-denominated variables are in 1990 PCE dollars.
\end{table}

\clearpage


\begin{table}[h]
\caption{Log Hourly Wage Equation Estimates}\label{tab:WageEqEst}
\begin{center}
\begin{tabular}{l c c}
\toprule
 & OLS & Selection-Bias Adjusted \\
\midrule
Age                         & $0.014^{***}$ & $0.007^{***}$ \\
                            & $(0.001)$     & $(0.002)$     \\
Education                   & $0.047^{***}$ & $0.041^{***}$ \\
                            & $(0.009)$     & $(0.009)$     \\
Black                       & $-0.092^{**}$ & $0.011$       \\
                            & $(0.037)$     & $(0.035)$     \\
Northeast                   & $0.206^{***}$ & $0.268^{***}$ \\
                            & $(0.079)$     & $(0.084)$     \\
Midwest                     & $0.087$       & $0.074$       \\
                            & $(0.059)$     & $(0.070)$     \\
West                        & $0.120$       & $0.184^{*}$   \\
                            & $(0.092)$     & $(0.097)$     \\
State Percent Services      & $0.016$       & $0.018^{*}$   \\
                            & $(0.010)$     & $(0.011)$     \\
State Percent Manufacturing & $0.004$       & $0.008$       \\
                            & $(0.007)$     & $(0.007)$     \\
State Percent Urban         & $0.003$       & $0.004$       \\
                            & $(0.003)$     & $(0.003)$     \\
Constant                    & $-0.024$      & $0.351$       \\
                            & $(0.333)$     & $(0.333)$     \\
\midrule
Obs                         & $1818$    & $3258$    \\
\bottomrule
\end{tabular}
\end{center}
\end{table}

\begin{table}[H]
	\vspace*{-0.7cm}
	\noindent\footnotesize * $(p<0.1)$, ** $(p<0.05)$, *** $(p<0.01)$. \\
	\emph{Notes:} Standard errors in parentheses are generated from a block bootstrap with replacement at the state level using 500 samples. The second column controls for a Heckman lambda based on a first stage probit which includes all the variables listed in the table and family size, the number of children under 6, the Food Stamp guarantee, the state unemployment rate, $N$, $G$, and $t$. 
\end{table}
\clearpage


\begin{table}[h]
\caption{First Stage Probit Instrument Comparison}
\begin{center}
\begin{footnotesize}
\begin{tabular}{l c c}
\toprule
 & Inverse Variance Weighted Average & Simple Average \\
\midrule
Log $\hat{W}$            & $-2.86^{***}$ & $-2.90^{***}$ \\
                         & $(0.78)$      & $(0.76)$      \\
Log (N+10)               & $-0.44^{***}$ & $-0.44^{***}$ \\
                         & $(0.04)$      & $(0.04)$      \\
Log G                    & $0.90^{***}$  & $0.93^{***}$  \\
                         & $(0.28)$      & $(0.29)$      \\
Log $\hat{W}(1-t)$       & $1.10^{*}$    & $1.11^{*}$    \\
                         & $(0.59)$      & $(0.59)$      \\
Age                      & $0.01$        & $0.01$        \\
                         & $(0.01)$      & $(0.01)$      \\
Black                    & $0.14$        & $0.14$        \\
                         & $(0.09)$      & $(0.09)$      \\
Family Size              & $-0.05$       & $-0.06$       \\
                         & $(0.05)$      & $(0.05)$      \\
Number of Children $<$ 6 & $0.30^{***}$  & $0.30^{***}$  \\
                         & $(0.04)$      & $(0.04)$      \\
Food Stamp Guarantee     & $2.01^{*}$    & $2.00$        \\
                         & $(1.22)$      & $(1.24)$      \\
Unemployment Rate        & $0.04^{*}$    & $0.04^{*}$    \\
                         & $(0.02)$      & $(0.02)$      \\
Northeast                & $0.34$        & $0.30$        \\
                         & $(0.29)$      & $(0.30)$      \\
Midwest                  & $0.15$        & $0.13$        \\
                         & $(0.26)$      & $(0.26)$      \\
West                     & $0.12$        & $0.11$        \\
                         & $(0.29)$      & $(0.29)$      \\
Log Z                    & $-0.25$       & $-0.22$       \\
                         & $(0.37)$      & $(0.30)$      \\
Constant                 & $4.74^{***}$  & $4.83^{***}$  \\
                         & $(1.78)$      & $(1.70)$      \\
\midrule
Obs                      & $3381$     & $3381$     \\
\bottomrule
\end{tabular}
\end{footnotesize}
\label{tab:FSPEstDetail}
\end{center}
\end{table}

\begin{table}[H]
	\vspace*{-0.7cm}
	\noindent\scriptsize * $(p<0.1)$, ** $(p<0.05)$, *** $(p<0.01)$. \\ Standard errors in parentheses are generated from a block bootstrap with replacement at the state level using 500 samples.
\end{table}

\begin{table}[!h]
\caption{\label{tab:LogZCorrMat}Pairwise Correlations Between Administrative Barrier Aggregates and Analysis Variables}
\begin{center}
\begin{tabular}[t]{lcc}
\toprule
 & Inv. Var. Wtg. & Simple Avg.\\
\midrule
Log $\hat{W}$ & $-0.298^{***}$ & $-0.366^{***}$\\
Log (N+10) & $-0.003^{\text{   }}$ & $-0.003^{\text{   }}$\\
Log G & $-0.404^{***}$ & $-0.337^{***}$\\
Log $\hat{W}$(1-t) & $0.173^{***}$ & $0.110^{***}$\\
Age & $-0.036^{*\text{  }}*$ & $-0.069^{***}$\\
Black & $0.070^{***}$ & $-0.001^{\text{   }}$\\
Family size & $0.023^{\text{   }}$ & $0.024^{\text{   }}$\\
No. Children $<$ 6 & $0.021^{\text{   }}$ & $0.029^{*\text{  }}$\\
Food Stamp Guarantee & $0.009^{\text{   }}$ & $-0.001^{\text{   }}$\\
Unemployment rate & $0.274^{***}$ & $0.287^{***}$\\
Northeast & $-0.384^{***}$ & $-0.587^{***}$\\
Midwest & $-0.009^{\text{   }}$ & $0.018^{\text{   }}$\\
West & $-0.040^{**\text{ }}$ & $0.161^{***}$\\
\bottomrule
\end{tabular}
\end{center}
\footnotesize * $(p<0.1)$, ** $(p<0.05)$, *** $(p<0.01)$ \\ \emph{Notes:} This table reports the pairwise correlation coefficients between aggregates of the log of the AFDC administrative barriers and the variables used in our empirical analysis.
\end{table}

\begin{table}[!h]
\caption{\label{tab:OPHIBal}Balance Given the Generalized Propensity Score: t-tests for Equality of Means}
\begin{center}
\resizebox{\linewidth}{!}{
\begin{tabular}[t]{llrrrrrrrr}
\toprule
\multicolumn{2}{c}{ } & \multicolumn{2}{c}{Specification 10} & \multicolumn{2}{c}{Specification 11} & \multicolumn{2}{c}{Specification 15} & \multicolumn{2}{c}{Specification 16} \\
\cmidrule(l{3pt}r{3pt}){3-4} \cmidrule(l{3pt}r{3pt}){5-6} \cmidrule(l{3pt}r{3pt}){7-8} \cmidrule(l{3pt}r{3pt}){9-10}
Log Z & Variable & Unadj. & GPS Adj. & Unadj. & GPS Adj. & Unadj. & GPS Adj. & Unadj. & GPS Adj.\\
\midrule
 & Log $\hat{W}$ & -15.24 & -0.51 & -16.25 & -1.50 & -16.96 & -0.33 & -18.29 & -2.63\\
 & Log (N+10) & -0.65 & 1.57 & -1.01 & 0.97 & -0.03 & 2.11 & 0.02 & 0.97\\
 & Log G & -8.78 & -0.73 & -9.39 & -0.60 & -8.76 & 1.38 & -15.83 & -1.01\\
 & Log $\hat{W}$(1-t) & 2.71 & 3.19 & 1.24 & 1.62 & 1.52 & 1.05 & 4.03 & 2.49\\
 & Age & -2.68 & 1.32 & -3.48 & 0.72 & -4.65 & 1.49 & -3.57 & -1.09\\
 & Black & 0.94 & -0.50 & 2.92 & -1.07 & 1.15 & -0.36 & 5.08 & -1.18\\
 & Number of Children $<$ 6 & 1.28 & -0.81 & 1.55 & -0.07 & 3.19 & -0.92 & 1.67 & 1.44\\
 & Family Size & 2.08 & 1.63 & 1.42 & 1.24 & 3.53 & 0.84 & 0.31 & 0.33\\
 & Food Stamp Guarantee & -0.92 & -0.09 & -0.59 & 0.03 & -0.27 & -1.86 & -1.47 & -1.98\\
 & Unemployment Rate & 13.24 & -1.58 & 13.54 & -1.64 & 12.31 & -0.62 & 17.22 & 0.18\\
 & Northeast & -32.72 & -19.94 & -33.08 & -17.69 & -33.61 & -8.18 & -32.22 & -16.76\\
 & Midwest & -1.98 & 6.54 & -1.90 & 4.79 & -1.88 & 0.19 & -2.67 & 4.35\\
\multirow{-13}{*}{\raggedright\arraybackslash Inv. Var. Wtg. Avg.} & West & 14.11 & 2.22 & 14.12 & 3.58 & 12.93 & 2.27 & 9.70 & 4.45\\
\midrule
 & Log $\hat{W}$ & -12.42 & -3.87 & -12.00 & -3.37 & -11.83 & -2.27 & -12.48 & -1.04\\
 & Log (N+10) & -1.00 & -1.19 & -0.92 & -0.56 & -0.18 & -0.43 & -1.10 & 0.40\\
 & Log G & -12.35 & -1.59 & -13.60 & -3.10 & -10.48 & 1.29 & -11.35 & -1.28\\
 & Log $\hat{W}$(1-t) & 1.23 & -1.10 & 2.52 & 0.07 & 1.23 & 0.08 & 2.04 & 1.76\\
 & Age & -2.88 & -1.86 & -2.02 & -1.39 & -2.22 & 0.66 & -3.38 & -0.24\\
 & Black & 1.42 & 1.80 & 3.28 & 1.29 & 1.67 & 1.36 & 2.36 & -0.04\\
 & Number of Children $<$ 6 & 0.49 & 0.15 & -1.10 & -1.04 & 0.48 & 1.20 & 0.87 & -1.21\\
 & Family Size & -1.79 & -1.87 & -2.30 & -2.34 & -0.27 & 3.43 & -1.10 & -0.97\\
 & Food Stamp Guarantee & -1.97 & -0.66 & -2.12 & -1.61 & -1.42 & -1.33 & -1.71 & -0.19\\
 & Unemployment Rate & 11.94 & 2.58 & 12.59 & 2.94 & 12.04 & -0.59 & 10.15 & 1.33\\
 & Northeast & -22.33 & -5.57 & -21.47 & -5.48 & -21.53 & -6.51 & -23.08 & -7.70\\
 & Midwest & -2.26 & 1.47 & -3.47 & 0.93 & -2.28 & 2.07 & -4.05 & 1.69\\
\multirow{-13}{*}{\raggedright\arraybackslash Simple Avg.} & West & 10.43 & 0.20 & 9.49 & 0.10 & 10.51 & 4.24 & 12.30 & 1.73\\
\bottomrule
\end{tabular}
}
\end{center}
\footnotesize \emph{Notes:} This table reports the t-test statistics for the equality of means for observations above and below the median value of the aggregates of the log of the AFDC administrative barriers. The unadjusted columns report the test statistics that do not adjust for the generalized propensity score (GPS). The GPS adjusted columns implement the procedure from \cite{HirImb2004}. The GPS is generated from an ordered probit model. All specifications include the AFDC participation probit covariates and different polynomials and interactions. Specification (10) includes cubes of the continuous covariates and interactions with $\log[W(1-t)]$. Specification (11) includes squares of the continuous covariates and interactions with black and $\log[W(1-t)]$. Specification (15) is the same as (10) but adds interactions with black and $\log(G)$. Specification (16) includes the continuous covariates squared and interactions with black, the food stamp guarantee, and the state unemployment rate.
\end{table}

\clearpage

\begin{table}[!h]
\UseRawInputEncoding
\caption{\label{tab:pFSumLogZs}F-Statistics by Different Participation Probability Ranges for Individual Administrative Barriers}
\begin{center}
\resizebox{\linewidth}{!}{
\begin{tabular}[t]{cccccccc}
\toprule
Part. Prob. Range & \makecell[c]{Pct. Ineligible\\in Error} & \makecell[c]{Pct. Hearings\\Appeals Denied} & \makecell[c]{Pct. Cases\\Denied Non-Grant} & \makecell[c]{Pct. Apps\\Denied} & \makecell[c]{Pct. Apps\\Denied Proc.} & \makecell[c]{Error Rate in\\Payment Det.} & \makecell[c]{Error Rate\\Resulting in Underpay.}\\
\midrule
0.00--0.33 & 0.87 & 0.31 & 0.12 & -0.14 & 0.39 & -0.36 & -0.65\\
0.33--0.66 & 1.47 & 0.71 & 3.42 & 8.45 & 5.35 & 6.86 & 4.14\\
0.66--1.00 & 0.29 & 0.22 & 0.61 & 2.30 & 1.54 & 2.49 & 1.07\\
\midrule
0.00--0.25 & 0.54 & -0.08 & -0.60 & -0.38 & 0.00 & -0.95 & 0.12\\
0.25--0.50 & 0.85 & 1.59 & 3.03 & 5.24 & 3.94 & 5.82 & 2.08\\
0.50--0.75 & 0.94 & -0.62 & 1.31 & 4.75 & 2.64 & 2.94 & 1.94\\
0.75--1.00 & 0.30 & 0.34 & 0.41 & 1.00 & 0.70 & 1.17 & 0.41\\
\bottomrule
\end{tabular}
}
\end{center}
\footnotesize \emph{Notes:} This table reports the F-statistics within different participation probability ranges. To calculate the F-statistic within a specific range of $\hat{F}$, define $RSS(q)$ as the residual sum of squares, equal to the sum of $[P-\hat{F}]^2$ taken over all observations in the range. The F-stat is calculated as (1) the difference in $RSS(q)$ for the restricted model excluding the instruments and the unrestricted model $RSS(q)$ including the instruments divided by the d.o.f., divided by (2) the residual variance computed over all observations in the sample, using $\hat{F}$ from the restricted model. Participation probabilities come from the probit model which include $\log(W)$, $\log[W(1-t)]$, $\log(G)$, $\log(N+10)$ (because some observations have $N=0$), age, black, family size, the number of children less than 6, the state unemployment rate, three regional dummies (a fourth is omitted), and the Food Stamp guarantee. Column headings denote which log AFDC administrative barrier is used in the probit specification. All specifications include interactions with $N$, $G$, and $W(1-t)$.
\end{table}

\clearpage

\section{Additional Figures}\label{app:addfigs}

\begin{figure}[ht]	
	\caption{Marginal Labor Supply Curves for Different Hours Equation Specifications}\label{fig:MTEDiffH5k}
	\hspace*{-2.5cm}
	\centering
	\begin{minipage}{\linewidth}
		\includegraphics[scale=0.85]{"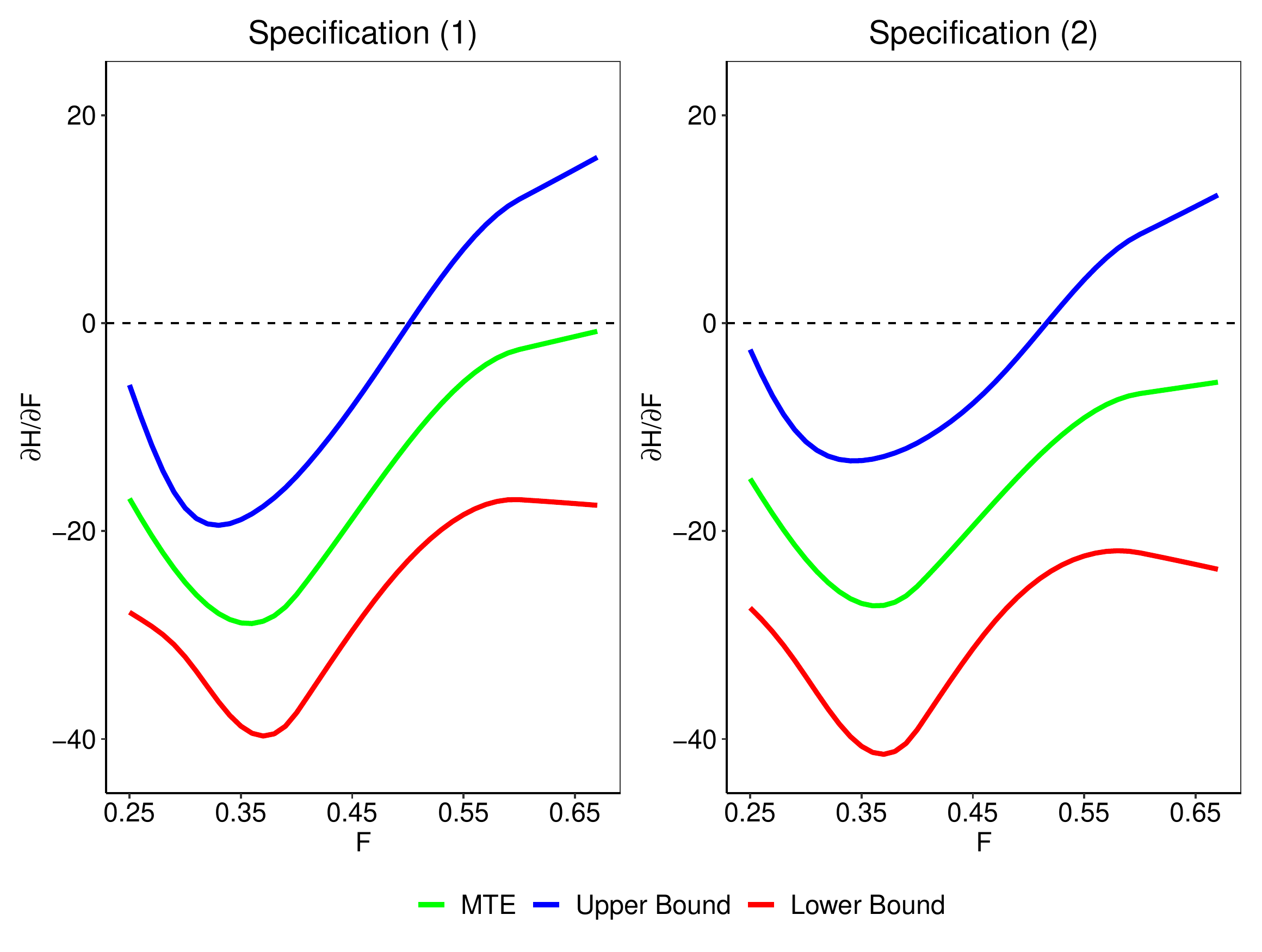"}
		\rule{\linewidth}{0in}
		\footnotesize
		\emph{Notes:} This figure plots the marginal treatment effect curves using a 5-knot cubic spline specification and a first stage probit model with the inverse variance weighted log of the AFDC administrative barriers and interactions with $N$, $G$, and $W(1-t)$. Each panel corresponds to specifications of the hours equation  found in Table \ref{tab:HoursEqSplineInt}. Upper and lower bounds are generated from a block bootstrap with replacement at the state level using 500 samples.
	\end{minipage}
\end{figure}

\begin{figure}[ht]	
	\caption{Marginal Labor Supply Curves Using Alternative 1992 Law Change Instrument}\label{fig:MTETimeSeriesAlt}
	\hspace*{-2.0cm}
	\centering
	\begin{minipage}{\linewidth}
		\includegraphics[scale=0.80]{"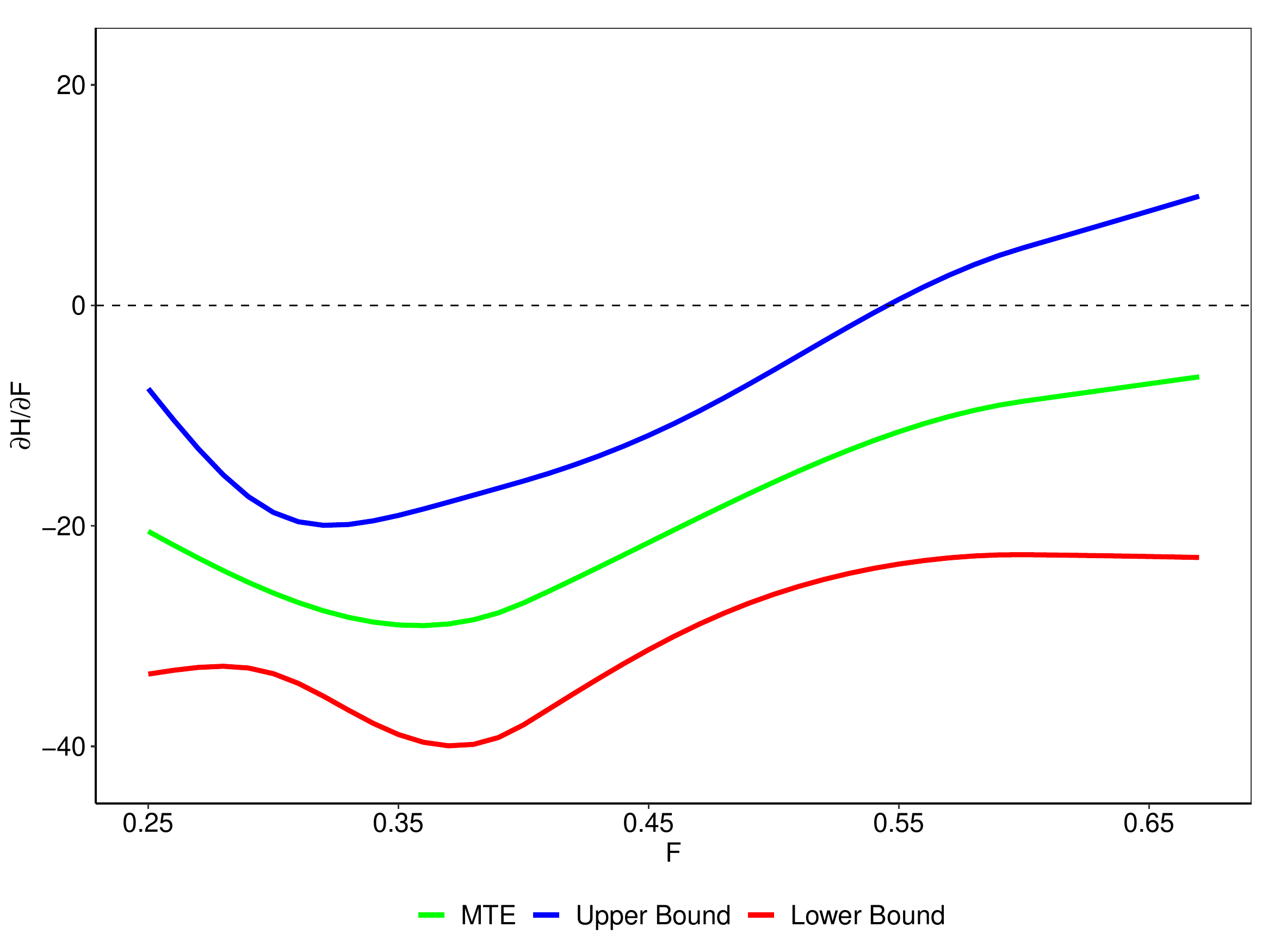"}
		\rule{\linewidth}{0in}
		\footnotesize
		\emph{Notes:} This figure plots the marginal treatment effect curves using a 5-knot cubic spline specification and the first stage probit model using a time trend and indicator for 1992 in column (2) of Table \ref{tab:LogZTSMicroReg}. Upper and lower bounds are generated from a block bootstrap with replacement at the state level using 500 samples.  
	\end{minipage}
\end{figure}
\clearpage

\section{Cubic Spline} \label{app:spline}
\doublespacing
The five-knot natural cubic spline is given here, using similar notation to \cite[p.~145]{Hastieetal2009}. Splines using different numbers of knots are analogous.  Let $F_{1}$, $F_{2}$, $F_{3}$, $F_{4}$, and $F_{5}$ denote the five knot points of $\hat{F}$, the predicted participation probability. The $g$ function is specified as \begin{equation}
	g(\hat{F})=g_{1}+g_{2}\hat{F}+g_{3}S3+g_{4}S4+g_{5}S5
\end{equation}
where
\begin{equation}
	S3=d_{1}-d_{4}
\end{equation}
\begin{equation}
	S4=d_{2}-d_{4}
\end{equation}
\begin{equation}
	S5=d_{3}-d_{4}
\end{equation}
where
\begin{equation}
	d_{1}=\frac{\max(0,\hat{F}-F_{1})-\max(0,\hat{F}-F_{5})}{F_{5}-F_{1}}
\end{equation}
\begin{equation}
	d_{2}=\frac{\max(0,\hat{F}-F_{2})-\max(0,\hat{F}-F_{5})}{F_{5}-F_{2}}
\end{equation}
\begin{equation}
	d_{3}=\frac{\max(0,\hat{F}-F_{3})-\max(0,\hat{F}-F_{5})}{F_{5}-F_{3}}
\end{equation}
\begin{equation}
	d_{4}=\frac{\max(0,\hat{F}-F_{4})-\max(0,\hat{F}-F_{5})}{F_{5}-F_{4}}
\end{equation}

\end{document}